\documentclass[aps,reprint,superscriptaddress,prb,longbibliography]{revtex4-2}
\usepackage[english]{babel}
\usepackage{amsmath,amssymb,mathtools,graphicx,epstopdf,tikz,tikz-3dplot,hyperref,svg}

\hypersetup{
    colorlinks=true,
    linkcolor=blue,
    citecolor=blue,      
    urlcolor=blue,
}

\newcommand{\cree}[2]{\hat{c}^{\dagger}_{\mathbf{#1},#2}}
\newcommand{\anii}[2]{\hat{c}_{\mathbf{#1},#2}}
\newcommand{\crea}[1]{\hat{a}^{\dagger}_{\mathbf{#1}}}
\newcommand{\ania}[1]{\hat{a}_{\mathbf{#1}}}
\newcommand{\A}[1]{\hat{A}_{\mathbf{#1}}}

\newcommand{\intbeta}[0]{\int_{0}^{\beta}}
\newcommand{\T}[0]{\hat{T}_{\tau}}
\newcommand{\Tbrak}[1]{\big\langle \T #1 \big\rangle}

\newcommand{\D}[2]{\mathcal{D}(\omega_\mathbf{#1},#2)}
\newcommand{\G}[2]{\mathcal{G}(\mathbf{#1},#2)}
\newcommand{\F}[2]{\mathcal{F}(\mathbf{#1},#2)}
\newcommand{\Fd}[2]{\mathcal{F}^{\dagger}(\mathbf{#1},#2)}

\newcommand{\Gn}[3]{\mathcal{G}^{(#1)}(\mathbf{#2},#3)}
\newcommand{\Fn}[3]{\mathcal{F}^{(#1)}(\mathbf{#2},#3)}
\newcommand{\Fdn}[3]{\mathcal{F}^{(#1)\dagger}(\mathbf{#2},#3)}
\newcommand{\da}[0]{\downarrow}
\newcommand{\ua}[0]{\uparrow}
\newcommand{\LO}[0]{\bar{\omega}}

\newcommand{\matssum}[1]{\sum\limits_{#1=-\infty}^{+\infty}}

\usetikzlibrary{positioning,decorations.markings,backgrounds,fit}

\newcommand{\DIAGRAM}[1]{\begin{tikzpicture}[baseline={(0,-0.1)},
edgenode/.style={draw=none,fill=none, inner sep = 0, minimum size=0cm},
vertexnode/.style={circle, draw=black, fill=black, inner sep = 0, minimum size=0.1cm},
]
\begin{scope}[very thick,decoration={markings,mark=at position 0.5 with {\arrow{>}}}]
#1
\end{scope}
\end{tikzpicture}}

\begin{document}
\title{Nonlinear electron-phonon interactions in Migdal-Eliashberg theory}
\author{Ingvar Zappacosta}
\email{ingvar.zappacosta@uantwerpen.be}
\affiliation{Theory of Quantum Systems and Complex Systems, Universiteit Antwerpen, B-2000 Antwerpen, Belgium}
\author{Matthew Houtput}
\affiliation{Theory of Quantum Systems and Complex Systems, Universiteit Antwerpen, B-2000 Antwerpen, Belgium}
\affiliation{Computational Material Physics, University of Vienna, Kolingasse 14-16, Vienna A-1090, Austria}
\author{Jacques Tempere}
\affiliation{Theory of Quantum Systems and Complex Systems, Universiteit Antwerpen, B-2000 Antwerpen, Belgium}

\begin{abstract}
Superconducting systems based on attractive electron-phonon interactions are the ones which are best understood at a fundamental level. They are well described using Eliashberg theory, which, unlike BCS theory, explicitly takes into account phonon dynamics. It is most often assumed that only linear electron-phonon interactions are relevant. However, for some superconductors like MgB$_2$ or hydride based superconductors, nonlinear electron-phonon interactions are known to contribute significantly, which is not taken into account in conventional Eliashberg theory. 
%Most commonly anharmonic phonon-phonon interactions are used to modify this theoretical framework, overlooking the fact that nonlinear electron-phonon coupling can also have significant contributions. 
We provide a modification to Eliashberg theory by introducing nonlinear electron-phonon interactions. We show that the Eliashberg equations remain unchanged apart from a nonlinear extension of the Eliashberg spectral function. This extended spectral function can be used as a baseline for future ab initio calculations. We use it to construct an analytical toy model and show that the nonlinear electron-phonon coupling affects the superconducting gap function on the imaginary and real axis and causes an increase in the superconducting critical temperature.

%Eliashberg theory is a theory of superconductivity which explicitly takes the pairing mechanism of Cooper pairs into account. The best understood superconducting systems are the ones that are realized by an attractive electron-phonon interaction. It is most often assumed that these electron-phonon interactions are only relevant up to first order in conventional Eliashberg theory. However, many superconductors, such as hydrides, have displayed signs of significant anharmonicity, causing a lack of theoretical descriptions that explain experimental data. Most commonly anharmonic phonon-phonon interactions are used to modify this theoretical framework. It is not the only way in which anharmonicity can be taken into account. We provide a modification to Eliashberg theory by introducing anharmonic electron-phonon interactions in a second quantization model. We show that the Eliashberg equations remain unchanged apart from an anharmonic extension of the Eliashberg spectral function. This extended spectral function can be used as a baseline for future ab initio calculations. (In the dispersionless single-band approximation for the phonons,) We use it to construct an analytical toy model for the spectral function. (We solve the Eliashberg equations with the aforementioned spectral function, and show that the nonlinear electron-phonon interaction) causes an increase in the critical temperature and changes the behavior of the superconducting gap function on the imaginary and real axis.
\end{abstract}

\maketitle

\section{Introduction}
Superconductivity is a fascinating display of quantum effects acting at macroscopically observable scales through many-body interactions. Conventional theories such as BCS theory \cite{BCS} and Migdal-Eliashberg theory \cite{Eliashberg1960, Eliashberg1961, allen1983theory, carbotte1990properties} are able to describe many superconducting systems, where the latter allows for a more accurate description due to it explicitly taking retardation effects into account. The retardation effects originate from the Cooper pairs inside of these superconducting systems, which are bound together through an attracting force coming from the electron-phonon coupling. Even though not all superconducting mechanisms are thought to involve phonon mediated interactions \cite{Stewart2017, Maksimov_2000, Boeri_2022, kresin-1993, lozovik-1976, Han2024}, it is the mechanism that is best understood at a fundamental level \cite{Stewart2017, Maksimov_2000, Boeri_2022, kresin-1993}. 

In 1960 Eliashberg formulated a theory of superconductivity \cite{Eliashberg1960, Eliashberg1961}, now called Eliashberg theory (or Migdal-Eliashberg theory within the approximations of Migdal \cite{Migdal}). Most superconductors which seem to be based on electron-phonon interactions are described well by this theory, but there are notable exceptions. For example, strong electron-phonon coupling which invalidates Migdal's approximation \cite{Schrodi2021, Alexandrov_2001, Hague_2007, hague-2008, Miller1998, Durajski2016}, presence of significant anharmonicity \cite{Mahan1993, Crespi-Cohen1993, Errea2013, Errea2014, Errea2015, Setty2021, Setty2024anharmonic, Galbaatar1991isotope, Lucrezi2023, Choi, bianco2023, Liu2001, Adolphs_2013} and other special cases \cite{Maksimov_2000, chubukov2020eliashberg, Dynes1986, tikhonov2020microwave, vanderMarel2019, Collignon2019, GASTIASORO2020}. Recently, a functional-integral approach has been constructed for derivation of the Eliashberg equations \cite{Protter2021, Dalal2023, Aase2023}. This approach appears to be more rigorous than a diagrammatic expansion, allowing for a systemic way to derive responses in the presence of Gaussian fluctuations \cite{Protter2021}, and might solve issues with the Migdal approximation or strong Coulomb repulsion \cite{Dalal2023} for example.

%Ever since, there have been discoveries of many superconductors which seem to be based on electron-phonon superconducting mechanisms but are not described well {tikhonov2020microwave, chubukov2020eliashberg, baquero1988eliashberg, Choi, Errea2013, Mahan1993, Errea2013, Crespi-Cohen1993, Setty2021, Setty2024anharmonic, Galbaatar1991isotope, Lucrezi2023, Choi, Liu2001, GASTIASORO2020, Collignon2019, vanderMarel2019, Errea2015, Hague_2007, Alexandrov_2001, Schrodi2021}. One class of superconductors is those that appear to be significantly affected by anharmonicity. Indeed, conventional Migdal-Eliashberg theory is based on a harmonic approximation for the mediating phonons and for the electron-phonon coupling.

Conventional Migdal-Eliashberg theory is based on a harmonic approximation for both the mediating phonons and for the electron-phonon coupling. A substantial number of possible expansions of this theory have been proposed to try to solve the anharmonicity problem, most often by including anharmonic phonon-phonon interactions up to arbitrary order \cite{Mahan1993, Errea2013, Errea2014, Errea2015, Crespi-Cohen1993, Setty2021, Setty2024anharmonic, Galbaatar1991isotope, Lucrezi2023, Choi, bianco2023}. These anharmonic phonon interactions renormalize the phonon energies and modify the isotope effect, mostly lowering the predicted critical temperature. This method has already been used to improve the conventional harmonic theory by, for example, accurately explaining the critical superconducting temperature of MgB$_2$ \cite{Choi} or improving a severe overestimation of the critical temperature for palladium hydrides \cite{Errea2013} in a harmonic theory. The latter is especially interesting: even though calculations with anharmonic phonon-phonon interactions better agree with experimental data, the suppression of $T_c$ now causes an underestimation.

``Anharmonicity'' is a broad term which can be captured in many different manners. Inclusion of processes with interaction vertices describing scattering between the electron and multiple phonons is another example. These kinds of nonlinear terms are different in nature than purely anharmonic phonon-phonon interactions, and they lead to different effects, such as increasing the critical temperature. Materials which display significant signs of anharmonicity might benefit from both a nonlinear phonon and electron-phonon description. Addition of the latter on top of the former in a theoretical framework could be crucial to explain the critical temperature of anharmonic materials better \cite{Liu2001, bianco2023}.

In this work, we extend the Hamiltonian as used in \cite{Marsiglio} by including the nonlinear 1-electron-2-phonon interaction to calculate the changes on the Eliashberg equations. We show that the Eliashberg equations are unaffected by this nonlinear electron-phonon term, except for a redefinition of the Eliashberg spectral function as first shown by \cite{Karakozov-Maksimov}. In Sec. \ref{sec:Anharmonic_theory} the derivation of the standard and real-axis Eliashberg equations is given in presence of nonlinearity. Using these results the nonlinear Eliashberg spectral function is obtained in Sec. \ref{subsec:Eliashberg_function}. In Sec. \ref{sec:Numerics} we construct a model Eliashberg spectral function. Use of this model shows that the critical temperature of the superconducting system is increased monotonically with the strength of the nonlinear coupling. It is shown that the gap function evaluated on the imaginary axis is renormalized, and that the gap and electron normalization function evaluated on the real axis are altered significantly in normalization and structure by nonlinearity.

\section{Nonlinear Migdal-Eliashberg theory} \label{sec:Anharmonic_theory}
\subsection{The nonlinear Eliashberg equations}
Describing superconductivity with retardation effects, including the nonlinear 1-electron-2-phonon interaction, begins with the following Hamiltonian\begin{align} \label{eq:Hamiltonian}
    \hat{H}=\;&\sum_{\mathbf{k}}\sum_{\sigma}\varepsilon_{\mathbf{k}}\cree{k}{\sigma}\anii{k}{\sigma}+\sum_{\mathbf{q}}\hbar\omega_{\mathbf{q}}\crea{q}\ania{q}\nonumber\\
    &+\frac{1}{\sqrt{\nu}}\sum_{\mathbf{k},\mathbf{k'}}\sum_{\sigma}g(\mathbf{k,k'})\A{k'-k}\cree{k'}{\sigma}\anii{k}{\sigma}\nonumber\\
    &+\frac{1}{2\nu}\sum_{\mathbf{k},\mathbf{k'},\mathbf{q}}\sum_{\sigma}\gamma(\mathbf{k,k',-q})\A{k'}\A{-q}\cree{k+k'-q}{\sigma}\anii{k}{\sigma}\nonumber\\
    &+\frac{1}{2\nu}\sum_{\mathbf{k},\mathbf{k'},\mathbf{q}}\sum_{\sigma,\sigma'}u(\mathbf{q})\cree{k+q}{\sigma}\cree{k'-q}{\sigma'}\anii{k'}{\sigma'}\anii{k}{\sigma}.
\end{align} 
Here, $\cree{k}{\sigma}$ and $\anii{k}{\sigma}$ create or annihilate an electron with momentum $\mathbf{k}$ and spin $\sigma$, with $\varepsilon_\mathbf{k}$ representing the electronic band structure. The operators $\crea{q}$ and $\ania{q}$ create or annihilate a phonon with wave number $\mathbf{q}$, where $\omega_\mathbf{q}$ is the phonon dispersion. For sake of simplicity we assume only a single phonon branch. The derivation generalizes straightforwardly to multiple branches by adding a branch index for each phonon frequency in the derivation, which should be summed over to include all branches. The operator $\A{q}=\ania{q}+\crea{-q}$ represents the fact that interactions with phonons can either be of absorptive or emissive nature. The normalization $\nu$ describes the number of lattice sites in the system, which will eventually be assumed large. The prefactors $g(\mathbf{k,k'})$, $\gamma(\mathbf{k,k',-q})$ and $u(\mathbf{q})$ are the matrix elements of the linear electron-phonon, the nonlinear electron-phonon and the Coulomb interaction, respectively. These matrix elements will be kept as general as possible for the derivation of the Eliashberg equations. 

To be able to set up the Eliashberg equations, one has to start by constructing a coupled pair of Dyson series for the superconducting system. First, one has to define the relevant Matsubara Green's functions or propagators for the theory\begin{align}
    \G{p}{\tau-\tau'}&=-\Tbrak{\anii{p}{\sigma}(\tau)\cree{p}{\sigma}(\tau')},\label{eq:G}\\
    \F{p}{\tau-\tau'}&=\Tbrak{\anii{-p}{\da}(\tau)\anii{p}{\ua}(\tau')},\label{eq:F}\\
    \Fd{p}{\tau-\tau'}&=\Tbrak{\cree{p}{\ua}(\tau)\cree{-p}{\da}(\tau')}.\label{eq:Fd}
\end{align}
The upper propagator is the standard free electron propagator, the other two are the anomalous propagators written in imaginary time. The latter two propagators are called anomalous since expectation values of this kind normally vanish with conservation of particle number. However, in the superconducting state they are assumed to be non-vanishing. The anomalous propagators are related to the order parameter of the Cooper pair condensate \cite{Mahan}. Derivation of the Eliashberg equations results in two identical self-consistent equations for the anomalous propagators. Therefore, the anomalous propagators are equal \cite{Mahan} up to a phase factor \cite{marsiglio2001, pavarini-2013}. In this paper we assume both anomalous propagators to be equal from the start, neglecting the potential phase shift. Lastly, the phonon propagator is defined as\begin{align}
    \D{q}{\tau-\tau'}&=-\Tbrak{\A{q}(\tau)\A{-q}(\tau')}.
\end{align}
Our model includes anharmonicity (nonlinearity) for the electron-phonon interaction but not for phonon-phonon interactions. Phonon anharmonicity can be added by adding a phonon-phonon interaction term to the Hamiltonian. The phonon self-interaction term introduces a Dyson series for the phonon propagator. This alters only the phonon energies and does not fundamentally affect the way how the 1-electron-2-phonon interaction is included in Eliashberg theory. If the phonon energies are extracted from experimental data, this phonon renormalization does not matter, since nature already includes all possible phonon-phonon interactions. If the reader has a preference for a theoretical description without any experimental input then the interaction could be included, or they could use ab initio calculations to find the renormalized phonon energies using the non-perturbative stochastic self-consistent harmonic approximation (SSCHA) \cite{Monacelli_2021}.

The interacting part of the Hamiltonian (\ref{eq:Hamiltonian}) can be used to perform an $S$-matrix expansion on the relevant propagators \cite{Mahan}\begin{align}
    \G{p}{\tau}&=-\Tbrak{\anii{p}{\sigma}(\tau)\hat{S}\cree{p}{\sigma}(0)},\label{eq:G_corr}\\
    \Fd{p}{\tau}&=\Tbrak{\cree{p}{\ua}(\tau)\hat{S}\cree{-p}{\da}(0)},\label{eq:F_corr}
\end{align}
with the $S$-matrix defined as\begin{align}
    \hat{S}=\sum_{n=0}^{+\infty}\frac{(-1)^n}{n!}\hat{T}_\tau\prod_{j=0}^{n}\left[\intbeta d\tau_j\hat{V}(\tau_j)\right]
\end{align}
and where the interacting part of the Hamiltonian\begin{align} \label{eq:H_int}
    \hat{V}=\hat{V}_g+\hat{V}_\gamma+\hat{V}_u
\end{align}
is the combined sum of the different interactions, which is the Hamiltonian (\ref{eq:Hamiltonian}) without the upper line. Each interaction is denoted with a subscript that represents its matrix element. 

A detailed explanation and derivation of the Dyson series can be followed in Appendix \ref{app:Dyson}. Here, we continue with a short discussion about the $S$-matrix expansion. The first and lowest order expansion of the propagators practically only result in corrections for the Coulomb interaction, since expectation values over an odd amount of phonon operators vanish. Technically, the 1-electron-2-phonon interaction also has a first order contribution. This can be shown to vanish as well and is discussed in Appendix \ref{app:Debye-Waller}. It is crucial that one does not neglect the contributions of the anomalous propagators in the Wick decomposition as first realized by Gor'kov \cite{Gor'kov}. The entire expression can then be transformed from imaginary time to imaginary frequency, and the anomalous propagators can be set equal. Then, the first order correction terms for the propagators can be promoted to a Dyson series. Following this reasoning one can find the coupled set of Dyson series\begin{align}
    \mathcal{G}(p)&=\mathcal{G}^{(0)}(p)\Big[1+\Sigma_u(p)\mathcal{G}(p)-\phi_u(p)\mathcal{F}(p)\Big],\label{eq:G_Dyson_C}\\
    \mathcal{F}^\dagger(p)&=\mathcal{G}^{(0)}(-p)\Big[\Sigma_u(-p)\mathcal{F}(p)+\phi_u(p)\mathcal{G}(p)\Big],\label{eq:F_Dyson_C}
\end{align}
where\begin{align}
    \Sigma_u(p)&=-\frac{1}{\beta}\matssum{n}\int \frac{d\mathbf{k}}{(2\pi)^3}\big|u(\mathbf{p-k})\big|^2\mathcal{G}(k),\label{eq:Sigma_u}
\end{align}
and\begin{align}
    \phi_u(p)&=-\frac{1}{\beta}\matssum{n}\int \frac{d\mathbf{k}}{(2\pi)^3}\big|u(\mathbf{p-k})\big|^2\mathcal{F}^\dagger(k)\label{eq:Omega_u}
\end{align}
are the Coulomb electron and anomalous self-energies. The expressions above make use of a shorter notation where $p=(\mathbf{p},ip_m)$ and $k=(\mathbf{k},ik_n)$. We also introduced the fermionic $p_m=(2m+1)\pi/\beta$ and bosonic $k_n=2n\pi/\beta$ Matsubara frequencies (where $m,n\in \mathbb{Z}$) and $\beta=1/(k_BT)$ with $k_B$ the Boltzmann constant and $T$ the temperature of the system.

\begin{figure}[ht]
    \centering
    \input{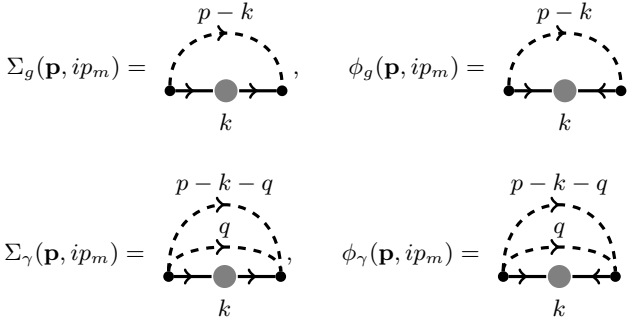}
    \caption{A visual representation of the self-energies which correct the propagators for the retarded interactions. The grey dot in the electron or anomalous propagators (full lines) denotes the fact that these are corrected propagators. The dashed lines are the phonon propagators.}
    \label{fig:self-energies}
\end{figure}

To include retardation effects in the superconductor, one has to perform at least a second order $S$-matrix expansion. This time, contributions of the Coulomb interaction, 1-electron-1-phonon and 1-electron-2-phonon exchange processes remain. The second order Coulomb correction can be neglected, following the reasoning that vertex corrections can be left out as an approximation \cite{Migdal}, combined with the fact that the promotion of the first order correction to a Dyson series includes the diagram with two (and infinitely more) consecutive Coulomb interactions. In Appendix \ref{app:Dyson} 1-electron-1-phonon and 1-electron-2-phonon corrections are promoted to Dyson series, yielding the same series as in the case of the Coulomb interaction (\ref{eq:G_Dyson_C}), (\ref{eq:F_Dyson_C}), but with different self-energies:\begin{widetext}\begin{gather}
    \Sigma_g(\mathbf{p},ip_m)=-\frac{1}{\beta}\matssum{n}\int \frac{d\mathbf{k}}{(2\pi)^3}\big|g(\mathbf{p},\mathbf{k})\big|^2\D{p-k}{ip_m-ik_n}\G{k}{ik_n},\label{eq:Sigma_g}\\
    \phi_g(\mathbf{p},ip_m)=-\frac{1}{\beta}\matssum{n}\int \frac{d\mathbf{k}}{(2\pi)^3}\big|g(\mathbf{p},\mathbf{k})\big|^2\D{p-k}{ip_m-ik_n}\Fd{k}{ik_n},\label{eq:Omega_g}\\
    \Sigma_\gamma(\mathbf{p},ip_m)=\frac{1}{2\beta^2}\matssum{n,l}\int\frac{d\mathbf{k}}{(2\pi)^3} \int \frac{d\mathbf{q}}{(2\pi)^3}\big|\gamma(\mathbf{p,p-k-q,q})\big|^2\D{p-k-q}{ip_m-ik_n-iq_l}\D{q}{iq_l}\G{k}{ik_n},\label{eq:Sigma_gamma}\\
    \hspace{-7.5px}\phi_\gamma(\mathbf{p},ip_m)=\frac{1}{2\beta^2}\matssum{n,l}\int\frac{d\mathbf{k}}{(2\pi)^3} \int \frac{d\mathbf{q}}{(2\pi)^3}\big|\gamma(\mathbf{p,p-k-q,q})\big|^2\D{p-k-q}{ip_m-ik_n-iq_l}\D{q}{iq_l}\Fd{k}{ik_n},\label{eq:Omega_gamma}
\end{gather}\end{widetext}
where $\D{q}{iq_l}=2\omega_\mathbf{q}/[(iq_l)^2-\omega_\mathbf{q}^2]$. The self-energies are shown in Figure \ref{fig:self-energies}. The corrections for electron-phonon exchange processes naturally include the phonon propagator in the self-energy, and two phonon propagators for the nonlinear coupling. These results are obtained by using that $g(\mathbf{k,k'})=g^\ast(\mathbf{k',k})$ and $\gamma(\mathbf{k,k',q})=\gamma^\ast(\mathbf{k+k'-q,q,k'})=\gamma^\ast(\mathbf{k+k'-q,-k',-q})$, the latter of which also allows us to impose $\gamma(\mathbf{k,k',q})=\gamma^\ast(\mathbf{k,q,k'})=\gamma^\ast(\mathbf{k,-k',-q})$. This is not an approximation or constraint on the Hamiltonian (\ref{eq:Hamiltonian}) since these conditions are already satisfied due to the Hamiltonian being hermitian. In addition, we do assume that $g(\mathbf{k,k'})=g^\ast(\mathbf{-k,-k'})$ and $\gamma(\mathbf{k,k',q})=\gamma(\mathbf{-k,k',q})$, meaning that the 1-electron-1-phonon vertex is invariant under time inversion and that the electron momentum can be reversed without affecting the 1-electron-2-phonon interaction. Moreover, our derivation assumes that tadpole diagrams do not contribute to the Dyson series.

\begin{figure}
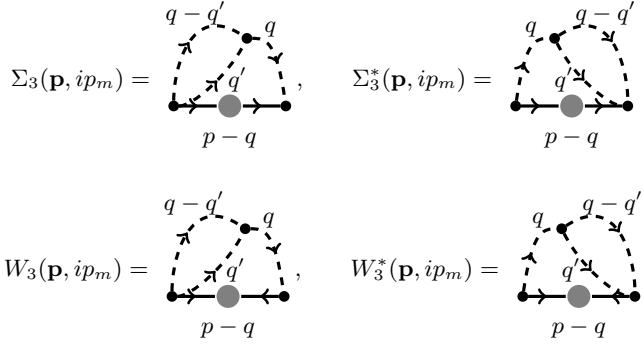

    \centering
    $\Sigma_3(\mathbf{p},ip_m)=$\DIAGRAM{
\node[vertexnode,opacity=0] (begin0) [below=0.3cm]{};
\node[vertexnode] (begin) [right=0.1cm of begin0]{};
\node[circle,fill=black,opacity=0.5,minimum size=0.3cm] (blob1) [right=0.5cm of begin]{}; \node[below=0cm of blob1]{$p-q$};
\node[vertexnode] (end1) [right=0.5cm of blob1]{};
\node[vertexnode] (3phonon) [left=0.375cm of end1, yshift=0.9cm]{};
\draw[black,postaction={decorate}] (begin) -- (blob1) {};
\draw[black,postaction={decorate}] (blob1) -- (end1) {};
\draw[dashed,postaction={decorate}] (3phonon) .. controls +(0.25,0) and +(-0.125,0.875) .. (end1) node[pos=0.5,above]{$q$};
\draw[dashed,postaction={decorate}] (begin) .. controls +(0,0.5) and +(-0.75,0.5) .. (3phonon) node[pos=0.6,above]{$q-q'$};
\draw[dashed,postaction={decorate}] (begin) .. controls +(0.25,0) and +(-0.25,-0.5) .. (3phonon) node[pos=0.6,right]{$q'$};
}\:,\qquad
$\Sigma_3^\ast(\mathbf{p},ip_m)=$\DIAGRAM{
\node[vertexnode,opacity=0] (begin0) [below=0.3cm]{};
\node[vertexnode] (begin) [right=0.1cm of begin0]{};
\node[circle,fill=black,opacity=0.5,minimum size=0.3cm] (blob1) [right=0.5cm of begin]{}; \node[below=0cm of blob1]{$p-q$};
\node[vertexnode] (end1) [right=0.5cm of blob1]{};
\node[vertexnode] (3phonon) [right=0.375cm of begin, yshift=0.9cm]{};
\node[edgenode] (ghost) [below=0.5cm of begin]{};
\draw[black,postaction={decorate}] (begin) -- (blob1) {};
\draw[black,postaction={decorate}] (blob1) -- (end1) {};
\draw[dashed,postaction={decorate}] (begin) .. controls +(0.125,0.875) and +(-0.25,0) .. (3phonon) node[pos=0.5,above]{$q$};
\draw[dashed,postaction={decorate}] (3phonon) .. controls +(0.75,0.5) and +(0,0.5) .. (end1) node[pos=0.4,above]{$q-q'$};
\draw[dashed,postaction={decorate}] (3phonon) .. controls +(0.25,-0.5) and +(-0.25,0) .. (end1) node[pos=0.4,left]{$q'$};
}
\\
\vspace{0.33cm}
$W_3(\mathbf{p},ip_m)=$\DIAGRAM{
\node[vertexnode,opacity=0] (begin0) [below=0.3cm]{};
\node[vertexnode] (begin) [right=0.1cm of begin0]{};
\node[circle,fill=black,opacity=0.5,minimum size=0.3cm] (blob1) [right=0.5cm of begin]{}; \node[below=0cm of blob1]{$p-q$};
\node[vertexnode] (end1) [right=0.5cm of blob1]{};
\node[vertexnode] (3phonon) [left=0.375cm of end1, yshift=0.9cm]{};
\draw[black,postaction={decorate}] (begin) -- (blob1) {};
\draw[black,postaction={decorate}] (end1) -- (blob1) {};
\draw[dashed,postaction={decorate}] (3phonon) .. controls +(0.25,0) and +(-0.125,0.875) .. (end1) node[pos=0.5,above]{$q$};
\draw[dashed,postaction={decorate}] (begin) .. controls +(0,0.5) and +(-0.75,0.5) .. (3phonon) node[pos=0.6,above]{$q-q'$};
\draw[dashed,postaction={decorate}] (begin) .. controls +(0.25,0) and +(-0.25,-0.5) .. (3phonon) node[pos=0.6,right]{$q'$};
}\:,\qquad
$W_3^\ast(\mathbf{p},ip_m)=$\DIAGRAM{
\node[vertexnode,opacity=0] (begin0) [below=0.3cm]{};
\node[vertexnode] (begin) [right=0.1cm of begin0]{};
\node[circle,fill=black,opacity=0.5,minimum size=0.3cm] (blob1) [right=0.5cm of begin]{}; \node[below=0cm of blob1]{$p-q$};
\node[vertexnode] (end1) [right=0.5cm of blob1]{};
\node[vertexnode] (3phonon) [right=0.375cm of begin, yshift=0.9cm]{};
\node[edgenode] (ghost) [below=0.5cm of begin]{};
\draw[black,postaction={decorate}] (begin) -- (blob1) {};
\draw[black,postaction={decorate}] (end1) -- (blob1) {};
\draw[dashed,postaction={decorate}] (begin) .. controls +(0.125,0.875) and +(-0.25,0) .. (3phonon) node[pos=0.5,above]{$q$};
\draw[dashed,postaction={decorate}] (3phonon) .. controls +(0.75,0.5) and +(0,0.5) .. (end1) node[pos=0.4,above]{$q-q'$};
\draw[dashed,postaction={decorate}] (3phonon) .. controls +(0.25,-0.5) and +(-0.25,0) .. (end1) node[pos=0.4,left]{$q'$};
}
    \caption{Possible self-energy corrections to the propagators of Eliashberg theory if the 3-phonon vertex is included. These diagrams are not included since we do not include any phonon-phonon interactions. inclusion of these diagrams result in an Eliashberg spectral function which is dependent of the Matsubara phonon propagator, making the resulting equations untractable.}
    \label{fig:diagram3ph}
\end{figure}

Higher order expansion terms only result in vertex corrections or corrections which are already included by the first and second order expansion. Vertex corrections are neglected since this derivation is assumed to apply in the framework of the Migdal theorem. If phonon anharmonicity using the 3-phonon vertex is included, it allows us to correct the propagators for diagrams as shown in Figure \ref{fig:diagram3ph}. It is a possibility that these diagrams are of the same order of as the diagrams in Figure \ref{fig:self-energies}, without phonon anharmonicity. However, we have not included these diagrams since they result in an additional self-energy term, causing the Eliashberg spectral function to not be writeable in its standard form anymore. It will be dependent of the Matsubara phonon propagator itself and receives a new Matsubara frequency dependence. This affects the real-axis Eliashberg equations \ref{subsec:analytic_continuation}, making them untractable.

The Dyson series for the Coulomb and the retarded interactions can now be combined to correct the propagators for all possible interactions. The Dyson series are identical to (\ref{eq:G_Dyson_C}) and (\ref{eq:F_Dyson_C}), differing only in the self-energies. This leaves us with the usual Dyson series from which the Eliashberg equations are set up \cite{Mahan}\begin{gather}
    \mathcal{G}(p)=\mathcal{G}^{(0)}(p)\Big[1+S(p)\mathcal{G}(p)-W(p)\mathcal{F}^\dagger(p)\Big],\label{eq:Dyson_G}\\
    \mathcal{F}^\dagger(p)=\mathcal{G}^{(0)}(-p)\Big[S(-p)\mathcal{F}^\dagger(p)+W(p)\mathcal{G}(p)\Big],\label{eq:Dyson_F}
\end{gather}
where the total self-energies are defined as\begin{gather}
    S(p)=\Sigma_u(p)+\Sigma_g(p)+\Sigma_\gamma(p),\label{eq:S(p)}\\
    W(p)=\phi_u(p)+\phi_g(p)+\phi_\gamma(p).
\end{gather}
The self-energies are the same as in the harmonic case, with additional contributions $\Sigma_\gamma(p)$ and $\phi_\gamma(p)$ due to the nonlinear 1-electron-2-phonon interaction. This addition still allows one to define\begin{align}
    S(p)=-\frac{1}{\beta}\matssum{n}\int\frac{d\mathbf{k}}{(2\pi)^3}V_\text{eff}(\mathbf{p,k},ip_m-ik_n)\mathcal{G}(k),\label{eq:S(p)_V_eff}\\
    W(p)=-\frac{1}{\beta}\matssum{n}\int\frac{d\mathbf{k}}{(2\pi)^3}V_\text{eff}(\mathbf{p,k},ip_m-ik_n)\mathcal{F}^\dagger(k).\label{eq:W(p)_V_eff}
\end{align}
Because the anomalous propagators vanish in the normal state, $W(p)$ is interpreted as the order parameter. The information of the different interactions is described by an effective potential\begin{widetext}\begin{align} \label{eq:V_eff}
    V_\text{eff}(\mathbf{p,k},ip_m-ik_n)&=\big|u(\mathbf{p-k})\big|^2+\big|g(\mathbf{p,k})\big|^2\D{p-k}{ip_m-ik_n}\nonumber\\
    &-\frac{1}{2\beta}\matssum{l}\int\frac{d\mathbf{q}}{(2\pi)^3}\big|\gamma(\mathbf{p,p-k-q,q})\big|^2\D{p-k-q}{ip_m-ik_n-iq_l}\D{q}{iq_l}.
\end{align}
The effective potential is still written in a way such that the self-energies are easily recognized. It can be simplified further by performing the Matsubara summation over $iq_l$. Using contour integration over the entire complex plane we can perform the following Matsubara summation\begin{align}
    -\frac{1}{\beta}\matssum{l}\D{p-k-q}{ip_m-ik_n-iq_l}\D{q}{iq_l}=\left(1+n_B^{\mathbf{p-k-q}}+n_B^{\mathbf{q}}\right)\mathcal{D}(\omega_\mathbf{p-k-q}+\omega_\mathbf{q},ip_m-ik_n)\nonumber\\
    +\big|n_B^{\mathbf{p-k-q}}-n_B^{\mathbf{q}}\big|\mathcal{D}(|\omega_\mathbf{p-k-q}-\omega_\mathbf{q}|,ip_m-ik_n),
\end{align}
which can be inserted in the effective potential, resulting in\begin{align} \label{eq:V_eff_calc}
    V_\text{eff}(\mathbf{p,k},ip_m-ik_n)=&\;\big|u(\mathbf{p-k})\big|^2+\big|g(\mathbf{p,k})\big|^2\D{p-k}{ip_m-ik_n}\nonumber\\
    &+\frac{1}{2}\int\frac{d\mathbf{q}}{(2\pi)^3}\big|\gamma(\mathbf{p,p-k-q,q})\big|^2\left(1+n_B^{\mathbf{p-k-q}}+n_B^{\mathbf{q}}\right)\mathcal{D}(\omega_\mathbf{p-k-q}+\omega_\mathbf{q},ip_m-ik_n)\nonumber\\
    &+\frac{1}{2}\int\frac{d\mathbf{q}}{(2\pi)^3}\big|\gamma(\mathbf{p,p-k-q,q})\big|^2\big|n_B^{\mathbf{p-k-q}}-n_B^{\mathbf{q}}\big|\mathcal{D}(|\omega_\mathbf{p-k-q}-\omega_\mathbf{q}|,ip_m-ik_n).
\end{align}
The factors $n_B^\mathbf{q}=1/(\exp(\beta\omega_\mathbf{q})-1)$ are the Bose-Einstein distributions. The effective potential does not obtain any new dependence by including the nonlinear electron-phonon coupling, meaning that the Eliashberg equations can be constructed from the Dyson series in the same manner as in the harmonic case. Standard practice is to assume isotropy from this point \cite{Mahan}. This can only be done by leaving out the momentum dependence of the self-energies and the electron momentum dependence of the matrix elements. For sake of completeness, we will derive the anisotropic Eliashberg equations first, after which an isotropic approximation can be made. Then the isotropic approximation can be discussed in more detail. Because only the effective potential has changed in our model without gaining any new dependence, the derivation for the anisotropic Eliashberg equations follows \cite{Mahan, Allen, Margine_Giustino} and results in
\begin{gather}
    Z(\mathbf{p},ip_m)=1+\frac{\pi}{p_m\beta}\matssum{n}\int\frac{d\mathbf{k}}{(2\pi)^3}\frac{\gamma_\mathbf{k}k_nZ(\mathbf{k},ik_n)}{\sqrt{k_n^2Z^2(\mathbf{k},ik_n)+W^2(\mathbf{k},ik_n)}}\lambda_\mathbf{p,k}(ip_m-ik_n),\label{eq:Eliashberg_Z}\\
    W(\mathbf{p},ip_m)=\frac{\pi}{\beta}\matssum{n}\int\frac{d\mathbf{k}}{(2\pi)^3}\frac{\gamma_\mathbf{k}W(\mathbf{k},ik_n)}{\sqrt{k_n^2Z^2(\mathbf{k},ik_n)+W^2(\mathbf{k},ik_n)}}\Big[\lambda_\mathbf{p,k}(ip_m-ik_n)-u^\ast\theta(\omega_c-|k_n|)\Big].\label{eq:Eliashberg_W}
\end{gather}
\end{widetext}
For completeness, the literature derivation has been outlined in Appendix \ref{app:Anisotropic}. Here, $\gamma_\mathbf{k}=\delta(\varepsilon_\mathbf{k}-\epsilon_F)$ (with $\varepsilon_\mathbf{k}$ the electronic band structure and $\epsilon_F$ the Fermi energy). The normalization function $Z$ is defined as\begin{align}
    ip_m\left[1-Z(\mathbf{p},ip_m)\right]=\frac{1}{2}\left[S(\mathbf{p},ip_m)-S(\mathbf{p},-ip_m)\right].
\end{align}
The auxiliary function $\lambda_\mathbf{p,k}$ describes the electron-phonon coupling, and represents the dimensionless coupling. It can be defined as the phonon part of the effective potential (\ref{eq:V_eff_calc})\begin{align} \label{eq:lambda_p,k}
    V_\text{eff}(\mathbf{p,k},ip_m-ik_n)=\big|u(\mathbf{p-k})\big|^2-\lambda_\mathbf{p,k}(ip_m-ik_n).
\end{align}
The contributions of the Coulomb interaction vanish for the electron self-energy, but not for the anomalous self-energy. There, we have introduced the notation\begin{align} \label{eq:u_isotropic}
    u=\frac{1}{g^2(\epsilon_F)}\int\frac{d\mathbf{p}}{(2\pi)^3}\int\frac{d\mathbf{k}}{(2\pi)^3}\gamma_\mathbf{p}\gamma_\mathbf{k}\big|u(\mathbf{p-k})\big|^2
\end{align}
with $(2\pi)^{3}g(\epsilon_F)=\int d\mathbf{k}\gamma_\mathbf{k}$ the density of states at the Fermi level. This definition is actually an isotropic average over the anisotropic system, even though we are still working in an anisotropic theory. Notoriously, the Coulomb interaction in Eliashberg theory is difficult to include in a mathematically rigorous way \cite{gross2018coulomb}. The standard ``workaround'' is to introduce the parameter $u$. In anisotropic Eliashberg theory, one must assume the Coulomb force to act isotropically when defining $u$.

The derivation of the Eliashberg equations approximates the typical phonon energies in the system to be much smaller than the Fermi energy. This can be a valid argument for the phonon term in the equations (i.e. $\lambda$), however, the Coulomb repulsion present in the anomalous self-energy does not act on the phonon energy scale. It is necessary to include an energy cutoff to avoid divergencies in the Matsubara summation. Usually this cutoff is a large value $\epsilon$, physically corresponding to the maximum energy of the electron band. It is possible to rescale $u$ to $u^\ast$ and lower the cutoff energy to a more computationally tractable size $\omega_c$ (often around ten times the highest phonon energy)\begin{align}
    u^\ast=\frac{u}{1+u\ln(\epsilon/\omega_c)},
\end{align}
where $u^\ast$ is called the Coulomb pseudopotential. This parameter is conventionally given a value of around $0.1-0.2$, coming from the calculations of Morel and Anderson \cite{Morel-Anderson} for a number of different materials, or used as a parameter to match experimental data. 

\begin{figure}
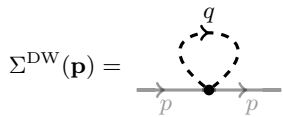

    \centering
    $\Sigma^\text{DW}(\mathbf{p})=$\hspace{-0.5cm}\DIAGRAM{
\node[vertexnode,opacity=0] (begin) [below=0.25cm]{};
\node[vertexnode] (blob1) [right=0.5cm of begin]{}; 
\node[vertexnode,opacity=0] (end1) [right=0.5cm of blob1]{};
\draw[black,postaction={decorate}, opacity = 0.4] (begin) .. controls +(-1,0) and +(0.33,0) .. (blob1) node[pos=0.5,below]{$p$};
\draw[black,postaction={decorate}, opacity = 0.4] (blob1) .. controls +(-0.33,0) and +(1,0) .. (end1) node[pos=0.5,below]{$p$};
\draw[dashed,postaction={decorate}] (blob1) .. controls +(-1.25,1) and +(1.25,1) .. (blob1) node[pos=0.5,above]{$q$};
}
    \caption{The Debye-Waller diagram \cite{Giustino2017}. The electron propagator is drawn for clarity,  but transparent, because it is not a part of the Debye-Waller self-energy. Since there is no intermediate electronic state, the self-energy is independent of any fermionic Matsubara frequency.}
    \label{fig:debye-waller}
\end{figure}

Inclusion of nonlinear electron-phonon coupling in the Hamiltonian actually introduces a Debye-Waller diagram, shown in Figure \ref{fig:debye-waller}. This diagram results in an additional contribution $\Sigma^\text{DW}(\mathbf{p})$ to the electron self-energy $S(p)$ (\ref{eq:S(p)}). Ultimately this does not affect the Eliashberg equations or spectral function, if assumed that\begin{align}
    \Sigma^\text{DW}(\mathbf{p})=-\frac{1}{2}\int\frac{d\mathbf{q}}{(2\pi)^3}\gamma(\mathbf{p,q,q})(1+2n_B^\mathbf{q})\ll \epsilon,
\end{align}
where $\epsilon$ is again the maximum energy of the electronic band. To explain briefly, for any given momentum $\mathbf{p}$, $\Sigma^\text{DW}(\mathbf{p})$ is a constant which has to be added only to the electron self-energy. Since it is independent of $ip_m$, it is symmetric in $ip_m$. Normally the symmetric part of $S(p)$ is identically 0 after the usual approximations \cite{Marsiglio}. Now this constant term remains. However, as long as this contribution is small compared to the energy cutoff $\epsilon$, it is negligible. The Debye-Waller diagram is treated more thoroughly in Appendix \ref{app:Debye-Waller}.

\subsection{The nonlinear Eliashberg function} \label{subsec:Eliashberg_function}
Equation (\ref{eq:lambda_p,k}) shows that the dimensionless coupling $\lambda_\mathbf{p,k}$ is directly linked to the phonon part of the effective potential, introduced in the self-energies of the electron and anomalous Dyson series. It is important to note that the dimensionless coupling can be written in a different, more well-known way\begin{align} \label{eq:lambda_p,k_Eliashberg}
    \lambda_\mathbf{p,k}(ip_m-ik_n)=-\int_0^{+\infty}d\nu\alpha^2F_\mathbf{p,k}(\nu)\mathcal{D}(\nu,ip_m-ik_n),
\end{align}
by definition of the anisotropic Eliashberg function $\alpha^2F_\mathbf{p,k}(\nu)$. Physically, the Eliashberg function tells us how dominant the electron-phonon interaction is for a specific frequency $\nu$, electron momentum $\mathbf{p}$ and phonon momentum $\mathbf{q}=\mathbf{p-k}$. Then, integrating this with the phonon propagator as weight, determines the average contribution of the electron-phonon coupling for a specific phonon mode $ip_m-ik_n$.

Rewriting the dimensionless coupling with the Eliashberg function has, to our knowledge, only been done for lowest order electron-phonon coupling. To include nonlinear terms it is important to notice that it is still possible to factor out the phonon propagator in the effective potential (\ref{eq:V_eff_calc}). An expression for the Eliashberg function can then be found by comparing the definition of the dimensionless coupling (\ref{eq:lambda_p,k}) with (\ref{eq:V_eff_calc}) and (\ref{eq:lambda_p,k_Eliashberg}), yielding\begin{widetext}\begin{align} \label{eq:Eliashberg_function_1branch}
    \alpha^2F_\mathbf{p,k}(\nu)&=\big|g(\mathbf{p,k})\big|^2B(\omega_\mathbf{p-k},\nu)+\int\frac{d\mathbf{q}}{(2\pi)^3}\frac{1+n_B^\mathbf{p-k-q}+n_B^\mathbf{q}}{2}\big|\gamma(\mathbf{p,p-k-q,q})\big|^2B(\omega_\mathbf{p-k-q}+\omega_\mathbf{q},\nu)\nonumber\\
    &+\int\frac{d\mathbf{q}}{(2\pi)^3}\frac{|n_B^\mathbf{p-k-q}-n_B^\mathbf{q}|}{2}\big|\gamma(\mathbf{p,p-k-q,q})\big|^2B(|\omega_\mathbf{p-k-q}-\omega_\mathbf{q}|,\nu).
\end{align}
This result can be generalized straightforwardly for multiple phonon branches by adding the branch indices $\alpha$ and $\beta$ to the phonon frequencies and matrix elements, and a summation over all branches\begin{align} \label{eq:Eliashberg_function}
    \alpha^2F_\mathbf{p,k}(\nu)&=\sum_{\alpha}\big|g_\alpha(\mathbf{p,k})\big|^2B(\omega_\mathbf{p-k}^{\alpha},\nu)+\sum_{\alpha,\beta}\int\frac{d\mathbf{q}}{(2\pi)^3}\frac{1+n_B^{\mathbf{p-k-q},\alpha}+n_B^{\mathbf{q},\beta}}{2}\big|\gamma_{\alpha\beta}(\mathbf{p,p-k-q,q})\big|^2B(\omega_{\mathbf{p-k-q}}^{\alpha}+\omega_\mathbf{q}^{\beta},\nu)\nonumber\\
    &+\sum_{\alpha,\beta}\int\frac{d\mathbf{q}}{(2\pi)^3}\frac{|n_B^{\mathbf{p-k-q},\alpha}-n_B^{\mathbf{q},\beta}|}{2}\big|\gamma_{\alpha\beta}(\mathbf{p,p-k-q,q})\big|^2B(|\omega_{\mathbf{p-k-q}}^{\alpha}-\omega_{\mathbf{q}}^{\beta}|,\nu)\Big].
\end{align}\end{widetext}
Here, we used the spectral function of the free phonon\begin{align} \label{eq:B_spectral}
    B(\omega_\mathbf{p-k},\nu)=\delta(\nu-\omega_\mathbf{p-k})-\delta(\nu+\omega_\mathbf{p-k}).
\end{align}
In theory the free propagator spectrum is usually seen as a set of infinitely sharp peaks. In reality these peaks are broadened due to other effects, such as anharmonic phonon-phonon interactions and general temperature effects.
The expression for the Eliashberg function shows the influence of the nonlinear electron-phonon coupling, which is now dependent of temperature through the Bose-Einstein distributions. The first term describes the harmonic part, where the Eliashberg function resembles the phonon spectrum with moderate changes due to the 1-electron-1-phonon matrix element possibly affecting the weight of the peaks. The next part, describing nonlinearity in the electron-phonon coupling is less obvious to interpret. At first sight it adds two other copies of the phonon spectrum to the Eliashberg function. However, the integration over the Brillouin zone smears out these contributions, once again with the relevant electron-phonon coupling matrix element as weight. The nonlinear part also increases for larger temperatures, due to the Bose-Einstein distributions in the prefactor. Therefore, superconducting systems with higher critical temperatures experience larger contributions from nonlinearity. Knowledge about the matrix element is key to be able to determine the full contribution of this nonlinear term. In section \ref{sec:Numerics} we will impose that the electron couples only to dispersionless phonons with a constant frequency $\LO$. This is a signature of Fröhlich theory \cite{Frohlich}, where dispersionless longitudinal optical phonons are considered to be the only contributions to polaron formation. In this case, it will allow us to determine some general contributions of nonlinear electron-phonon interactions without knowing the structure of the electron-phonon coupling.

\subsection{Analytic continuation} \label{subsec:analytic_continuation}
The critical temperature of the superconducting system can be determined from the standard Eliashberg equations. However, the imaginary-axis Matsubara formalism is insufficient to determine any dynamical property, such as optical response and many other properties \cite{Marsiglio}. Additionally, the self-energies evaluated on the imaginary axis are hard to interpret physically. Therefore, it is desired to perform an analytic continuation on the Eliashberg equations to understand superconducting systems better. Simply setting $ip_m\rightarrow\omega+i\delta$ in (\ref{eq:Eliashberg_Z}) and (\ref{eq:Eliashberg_W}) does not yield the correct result due to the presence of the Matsubara summation over $k_n$. The starting point are the expressions of the self-energies $S(p)$ (\ref{eq:S(p)_V_eff}) and $W(p)$ (\ref{eq:W(p)_V_eff}). 
%It is possible to perform the Matsubara summation after using the spectral representations of the Green's functions in the self-energy expressions. Once the summation is done, the analytic continuation on the self-energies can be done by changing the frequency dependence $ip_m\rightarrow\omega+i\delta$ from the imaginary axis to the real axis. After this the expressions can be simplified further to obtain the real-axis Eliashberg equations. 
The full derivation will not be given here, for it is explained entirely in \cite{Marsiglio_analytical}. The derivation still holds in the nonlinear case since the nonlinear contribution is only added to the effective potential without introducing any new dependence (\ref{eq:V_eff_calc}). Therefore, the real-axis Eliashberg equations are unchanged, given by \cite{Marsiglio, Margine_Giustino, Marsiglio_analytical}\begin{widetext}\begin{align}
    Z(\mathbf{p},\omega+i\delta)=1+\frac{\pi}{\omega\beta}\matssum{n}\int &\frac{d\mathbf{k}}{(2\pi)^3}\frac{\gamma_\mathbf{k}ik_nZ(\mathbf{k},ik_n)}{\sqrt{k_n^2Z^2(\mathbf{k},ik_n)+W^2(\mathbf{k},ik_n)}}\lambda_\mathbf{p,k}(\omega-ik_n)\nonumber\\
    +\frac{i\pi}{\omega}\int\frac{d\mathbf{k}}{(2\pi)^3}\int_0^{+\infty}dz\alpha^2F_\mathbf{p,k}(z)\bigg\{ &\left[n_B(z)+n_F(z+\omega)\right] \frac{\gamma_\mathbf{k}(\omega+z)Z(\mathbf{k},\omega+z+i\delta)}{\sqrt{(\omega+z)^2Z^2(\mathbf{k},\omega+z+i\delta)-W^2(\mathbf{k},\omega+z+i\delta)}}\nonumber\\
    +&\left[n_B(z)+n_F(z-\omega)\right] \frac{\gamma_\mathbf{k}(\omega-z)Z(\mathbf{k},\omega-z+i\delta)}{\sqrt{(\omega-z)^2Z^2(\mathbf{k},\omega-z+i\delta)-W^2(\mathbf{k},\omega-z+i\delta)}}\bigg\},\label{eq:real-axis_Z}\\
    W(\mathbf{p},\omega+i\delta)=\frac{\pi}{\beta}\matssum{n}\int\frac{d\mathbf{k}}{(2\pi)^3}&\frac{\gamma_\mathbf{k}W(\mathbf{k},ik_n)}{\sqrt{k_n^2Z^2(\mathbf{k},ik_n)+W^2(\mathbf{k},ik_n)}}\left[\lambda_\mathbf{p,k}(\omega-ik_n)-u^\ast\theta(\omega_c-|k_n|)\right]\nonumber\\
    +i\pi\int\frac{d\mathbf{k}}{(2\pi)^3}\int_0^{+\infty}dz\alpha^2F_\mathbf{p,k}(z)\bigg\{ &\left[n_B(z)+n_F(z+\omega)\right] \frac{\gamma_\mathbf{k}W(\mathbf{k},\omega+z+i\delta)}{\sqrt{(\omega+z)^2Z^2(\mathbf{k},\omega+z+i\delta)-W^2(\mathbf{k},\omega+z+i\delta)}}\nonumber\\
    +&\left[n_B(z)+n_F(z-\omega)\right] \frac{\gamma_\mathbf{k}W(\mathbf{k},\omega-z+i\delta)}{\sqrt{(\omega-z)^2Z^2(\mathbf{k},\omega-z+i\delta)-W^2(\mathbf{k},\omega-z+i\delta)}}\bigg\}.\label{eq:real-axis_W}
\end{align}\end{widetext}
Even though the equations are unaffected by nonlinearity, the Eliashberg function has a new definition (\ref{eq:Eliashberg_function}) compared to the case without 1-electron-2-phonon interactions, being dependent of temperature as well. To solve the real-axis equations, the self-energies on the imaginary axis have to be determined first, since they are required as an input in the top line of equations (\ref{eq:real-axis_Z}) and (\ref{eq:real-axis_W}). It is essential that the branch cut is fixed on the positive real axis for computation of the coupled self-consistent equations. This is only the case for square roots over the self-energies evaluated on the real-axis, since these are complex-valued functions, on the imaginary axis they are real. This condition is an artifact of the energy integral $\mathcal{E}$, where a complex-valued integrand requires contour integration. This integration has to be done over the upper-half complex plane since this is the domain where the retarded Green's functions are analytic.

\subsection{Isotropic approximation}
Up to this point we worked in an anisotropic theory. Computationally this can be very demanding since the self-energy equations are coupled for all momenta. For this reason it is often assumed that the system is isotropic. A general way of obtaining the isotropic equations is through averaging the anisotropic equations over the Fermi surface\begin{align}
    \int\frac{d\mathbf{p}}{(2\pi)^3}\frac{\gamma_\mathbf{p}}{g(\epsilon_F)}Z(\mathbf{p},ip_m)&=Z(ip_m)\\
    \int\frac{d\mathbf{p}}{(2\pi)^3}\frac{\gamma_\mathbf{p}}{g(\epsilon_F)}W(\mathbf{p},ip_m)&=W(ip_m).
\end{align}
Instead of simply assuming isotropy, the isotropic self-energies can be approximated with a more careful approach. The Eliashberg equations (\ref{eq:Eliashberg_Z}), (\ref{eq:Eliashberg_W}) in this case can be simplified to\begin{widetext}\begin{gather}
    Z(ip_m)=1+\frac{\pi}{p_m\beta}\matssum{n}\frac{k_nZ(ik_n)}{\sqrt{k_n^2Z^2(ik_n)+W^2(ik_n)}}\lambda(ip_m-ik_n),\label{eq:Z_isotropic}\\
    W(ip_m)=\frac{\pi}{\beta}\matssum{n}\frac{W(ik_n)}{\sqrt{k_n^2Z^2(ik_n)+W^2(ik_n)}}\left[\lambda(ip_m-ik_n)-u^\ast\theta(\omega_c-|k_n|)\right],\label{eq:W_isotropic}
\end{gather}\end{widetext}
where the isotropic dimensionless coupling is defined as\begin{align}
    \lambda(ip_m-ik_n)=\int\frac{d\mathbf{p}}{(2\pi)^3}\int\frac{d\mathbf{k}}{(2\pi)^3}\frac{\gamma_\mathbf{p}\gamma_\mathbf{k}}{g(\epsilon_F)}\lambda_\mathbf{p,k}(ip_m-ik_n).
\end{align}
Equation (\ref{eq:lambda_p,k_Eliashberg}) shows that all anisotropy in the electron-phonon coupling originates from the Eliashberg function, therefore\begin{align}
    \alpha^2F(\nu)=\int\frac{d\mathbf{p}}{(2\pi)^3}\int\frac{d\mathbf{k}}{(2\pi)^3}\frac{\gamma_\mathbf{p}\gamma_\mathbf{k}}{g(\epsilon_F)}\alpha^2F_\mathbf{p,k}(\nu).
\end{align}
The normalization function $Z$ and self-energy $W$ are determined in a self-consistent manner, isotropic or not. Therefore, only the Eliashberg function has to be averaged over the Fermi surface to go from an anisotropic theory to an isotropic one in practice. The isotropic approximation can be made for the real-axis Eliashberg equations (\ref{eq:real-axis_Z}), (\ref{eq:real-axis_W}) as well. This is done by leaving out all momentum dependencies, all momentum integrations and every factor $\gamma_\mathbf{k}$ by using the above two expressions. The resulting isotropic equations are equal to the harmonic case described in \cite{Marsiglio}.

\section{Analytical toy model} \label{sec:Numerics}
\begin{figure}
    \centering
    \includegraphics[scale=0.45]{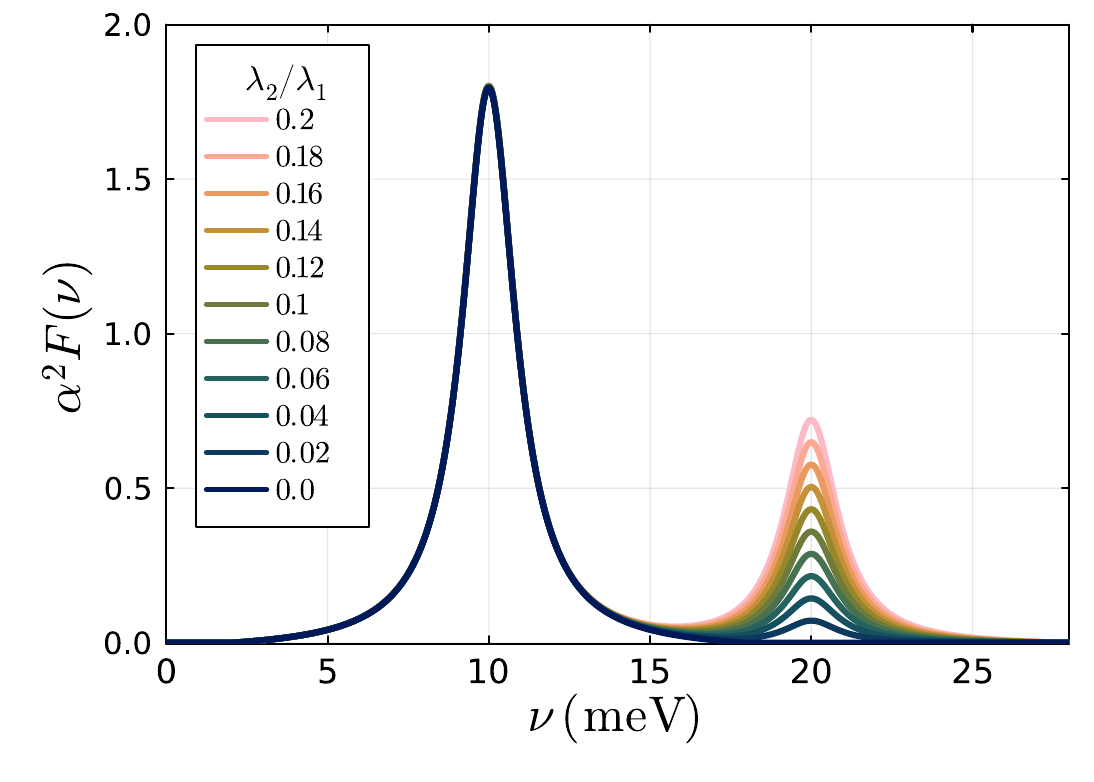}
    \caption{The Eliashberg function in the dispersionless phonon approximation with nonlinear electron-phonon coupling at $T=0$ K. Here we used $\lambda_1=1$ for many different $\lambda_2=\lambda_2(T=0)$ values, $\LO=10$ meV, $\delta=1$ meV and $\nu_c=8$ meV. One can clearly see the contributions of the harmonic and nonlinear parts, represented by the peaks at $\LO$ and $2\LO$ respectively. The model Eliashberg function has been rescaled so that (\ref{eq:integral_Eliashberg}) is satisfied. Notice that the peak at $2\LO$ is twice as high as the fraction of $\lambda_2/\lambda_1$ times the height of the peak at $\LO$, since the relation between the Eliashberg function and $\lambda$ goes like $\alpha^2F(\nu)/\nu$. This could also be seen in (\ref{eq:Eliashberg_LO}), where the phonon spectral function with weight $\lambda_2$ has an additional prefactor of 2.}
    \label{fig:Eliashberg_function}
\end{figure}

For computations, we will continue with the isotropic theory. The only input required for the Eliashberg equations is the Eliashberg spectral function $\alpha^2F(\nu)$, a value for the Coulomb parameter $u^\ast$ and its cutoff $\omega_c$. This manuscript focuses on the effect of nonlinear electron-phonon coupling. Therefore, we simply adopt a value of $u^\ast=0.12$ to include some Coulomb effects, a value which is typical for homogeneous metals \cite{Morel-Anderson}. The Coulomb cutoff is placed at 10 times the highest phonon energy in the spectrum. This choice of cutoff energy will be self-consistently justified from the results. The nonlinear Eliashberg function is given by (\ref{eq:Eliashberg_function}) and is dependent of the linear and nonlinear matrix elements and the spectral function of the free phonon. Computing the Eliashberg function for any real material is not a simple task, even less so with inclusion of nonlinearity. We impose that the electrons couple only to one branch of dispersionless phonons, with constant frequency $\LO$. Then, every phonon propagator or spectral function can be replaced by its dispersionless phonon counterpart, yielding\begin{widetext}\begin{align}
    \alpha^2F(\nu)=\int\frac{d\mathbf{q}}{(2\pi)^3}\gamma_\mathbf{q}\big|\overline{g}(\mathbf{q})\big|^2B(\LO,\nu)+\frac{1}{2}\int\frac{d\mathbf{q}}{(2\pi)^3}\int\frac{d\mathbf{q'}}{(2\pi)^3}\gamma_\mathbf{q}\big|\overline{\gamma}(\mathbf{q-q',q'})\big|^2\left[1+2n_B(\LO)\right]B(2\LO,\nu),
\end{align}\end{widetext}
with the isotropic matrix elements defined as\begin{gather}
    |\overline{g}(\mathbf{q})|^2=\int\frac{d\mathbf{p}}{(2\pi)^3}\frac{\gamma_\mathbf{p}}{g(\epsilon_F)}|g(\mathbf{p,p-q})|^2,\\
    |\overline{\gamma}(\mathbf{q-q',q'})|^2=\int\frac{d\mathbf{p}}{(2\pi)^3}\frac{\gamma_\mathbf{p}}{g(\epsilon_F)}|\gamma(\mathbf{p,p-q-q',q'})|^2.
\end{gather}
The main reason for this approximation is the fact that the constant phonon frequency allows us to decouple the matrix elements from the spectral functions. Therefore, computations using the spectral function can be performed without having done ab initio calculations about the phonon spectrum and the matrix elements. It is also an assumption made by \cite{houtput-2022}, where an explicit form of the 1-electron-2-phonon matrix element is presented in the dispersionless longitudinal optical phonon approximation with addition of some other approximations. Of course, the dispersionless phonons results in a simplified model which might not be applicable to real materials, but it does offer some intuition about the core effects of a nonlinear Eliashberg function. Performing the momentum integrations over the matrix elements results in some constant values\begin{align} \label{eq:Eliashberg_LO}
    \alpha^2F(\nu)= \frac{\LO}{2}\left[\lambda_1B(\LO,\nu)+2\lambda_2B(2\LO,\nu)\right],
\end{align}
where\begin{gather}
    \lambda_1=\frac{2}{\LO}\int\frac{d\mathbf{q}}{(2\pi)^3}\gamma_\mathbf{q}\big|\overline{g}(\mathbf{q})\big|^2,\\
    \lambda_2=\frac{1+2n_B(\LO)}{2\LO}\int\frac{d\mathbf{q}}{(2\pi)^3}\int\frac{d\mathbf{q'}}{(2\pi)^3}\gamma_\mathbf{q}\big|\overline{\gamma}(\mathbf{q-q',q'})\big|^2. \label{eq:lambda2(T)}
\end{gather}
Notice again that $\lambda_2$ is temperature dependent through $n_B(\LO)$. The weights of the spectral functions in (\ref{eq:Eliashberg_LO}) are chosen so that\begin{align} \label{eq:integral_Eliashberg}
    \lambda(0)=\int_0^{+\infty}\frac{2\alpha^2F(\nu)d\nu}{\nu}=\lambda_1+\lambda_2.
\end{align}
Instead of a phonon spectrum with delta peaks at the energy of the dispersionless phonons (\ref{eq:B_spectral}), we take into account broadening of the spectral lines by describing the spectral functions with Lorentzian profiles $B(\LO,\nu)\rightarrow B_\delta(\LO,\nu)$, being the typical shape of homogeneously broadened spectral lines. Adopting the same conventions as in \cite{Marsiglio} yields the following
\begin{align}
    &B_\delta(\LO,\nu) \nonumber \\
    &=\frac{1}{\pi}\left[\frac{\delta}{(\nu-\LO)^2+\delta^2}-\frac{\delta}{\nu_c^2+\delta^2}\right]\theta(|\nu-\LO|-\nu_c).
\end{align}
Here, $\theta(x)$ is the Heavyside step-function, which cuts off the Lorentzian at 0 for energies $\nu=\LO\pm\nu_c$, introducing a cutoff scale $\nu_c$. The Lorentzian expression for the spectral function implicitly assumes that $\nu>0$. This is indeed the domain we restrict our expressions to, but technically the spectral function has to be antisymmetric as in (\ref{eq:B_spectral}). Mathematically, this is a nascent delta function which converges weakly to the delta distribution\begin{align}
    \lim_{\delta\rightarrow0}B_\delta(\LO,\nu)=\delta(\nu-\LO)=B(\LO,\nu>0).
\end{align}
$\delta$ represents the broadening of the spectral lines, being the half-width at half-maximum. Setting $\delta\rightarrow0$ should indeed retrieve the original spectrum. If the spectral lines are broadened the integral in (\ref{eq:integral_Eliashberg}) decreases in value. To compensate for this we will renormalize the Eliashberg function $\alpha^2F(\nu)$ so that $\lambda(0)$ is kept constant under varying $\delta$. In Figure \ref{fig:Eliashberg_function} the Eliashberg function of this model is shown.

\subsection{Critical temperature}
The model Eliashberg function can be used to calculate some properties of the superconductor. First, the critical temperature $T_c$ is computed by using the standard Eliashberg equations, coupled for $Z$ (\ref{eq:Z_isotropic}) and $W$ (\ref{eq:W_isotropic}) on the imaginary axis. The temperature dependence in these equations comes from $\beta=1/(k_BT)$, the Matsubara frequencies and $\lambda_2(T)$. Since $W$ is the order parameter it only has a nonzero value in the superconducting state, $T<T_c$. So, to find $T_c$ one can assume $W$ to be infinitely small and calculate for which $T$ the coupled equations are satisfied. 

\begin{figure}
    \centering
    \includegraphics[scale=0.425]{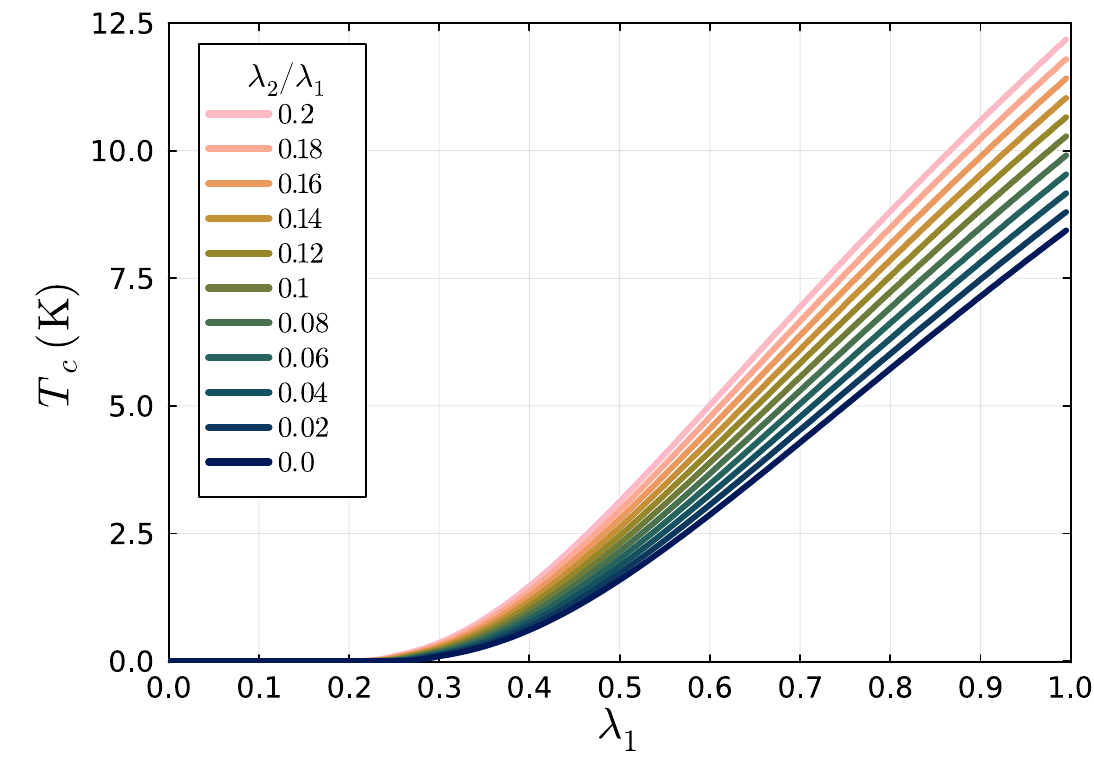}
    \caption{The critical temperature of the superconductor for a range of $\lambda_1$ and $\lambda_2=\lambda_2(T=0)$ values with $\LO=10$ meV, $\delta=1$ meV, $\nu_c=8$ meV, $u^\ast=0.12$ and $\omega_c=280$ meV (10 times the highest phonon energy in the spectrum). The critical temperature increases monotonically with $\lambda_1$ and $\lambda_2$. The essence is clear, inclusion of the 1-electron-2-phonon interaction strengthens the total electron-phonon coupling which raises the critical temperature.}
    \label{fig:superconducting_Tc}
\end{figure}

Figure \ref{fig:superconducting_Tc} shows the critical temperature for a range of $\lambda_1$ and $\lambda_2\coloneqq\lambda_2(T=0)$ values, which increases monotonically with $\lambda_1$ and $\lambda_2$. Looking at the Eliashberg spectral function (\ref{eq:Eliashberg_function}) it is clear that both the harmonic and nonlinear part are always positive quantities for $\nu\geq 0$. So, we indeed expect an increase of $T_c$ since the electron-phonon coupling binds the Cooper pairs together, which are bound together more strongly with this additional term in the electron-phonon coupling. The increase of the critical temperature can be significant for stronger nonlinear coupling strength. For example, setting $\lambda_2=0$ to $0.1$ for $\lambda_1=1$ raises $T_c$ from $8.5$ K to $10.4$ K, a relative increase of $22\%$. This trend continues for higher phonon energies. If the entire phonon spectrum is scaled up by a factor of 10, the critical temperature raises from $79.3$ K to $97.8$ K, the same relative increase and consistent with the asymptotic limit $T_c \sim\bar{\omega}\sqrt{\lambda}$. Increase in the temperature also affects $\lambda_2(T)$. However, the temperature regime in this model is too small to have a notable impact. For example, $\lambda_2(T=15\ $K$)/\lambda_2=1.00087$. Since there is currently no computationally viable method of calculating the full 1-electron-2-phonon matrix element $\gamma(\mathbf{p,k,q})$, it is yet unsure how large $\lambda_2$ can become for superconductors that display significant signs of anharmonicity. However, advances in ab initio calculations might give us the ability to compute these nonlinear matrix elements in the near future. In particular, a recent paper \cite{houtput2024} reporting a semi-analytical expression of the long-range part of this matrix element where it is linked directly to microscopic quantities of the material.

\subsection{The imaginary axis gap function}
\begin{figure}
    \centering
    \includegraphics[scale=0.425]{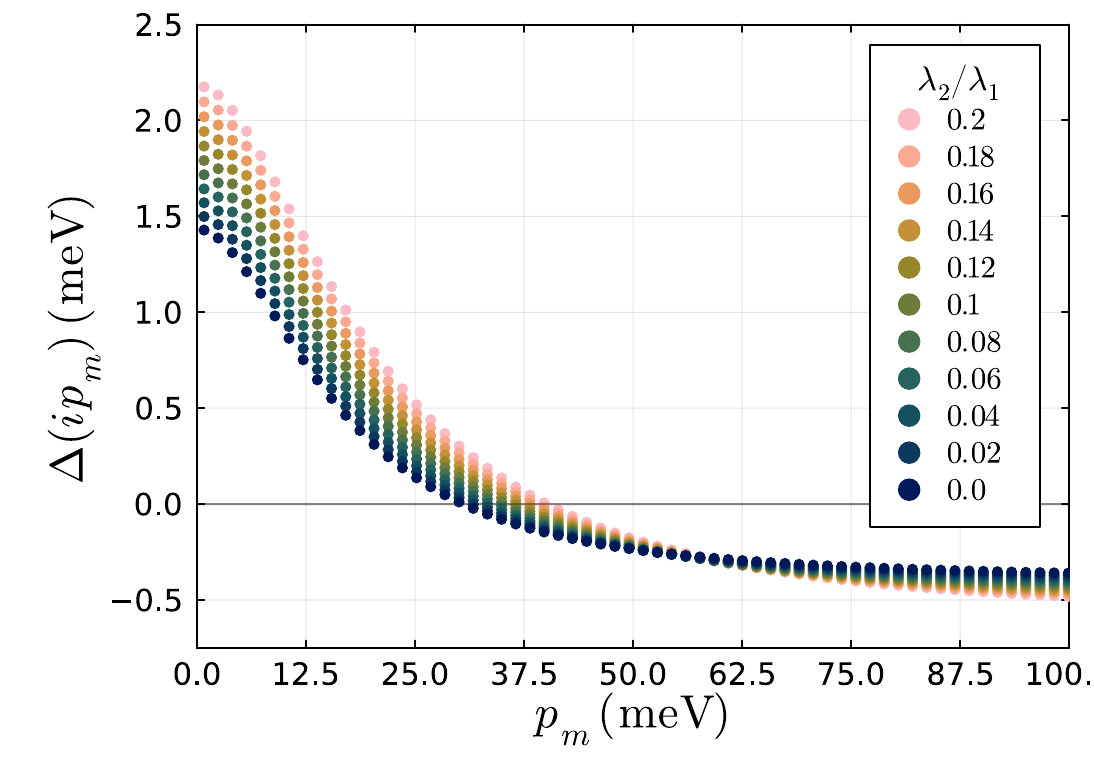}
    \includegraphics[scale=0.425]{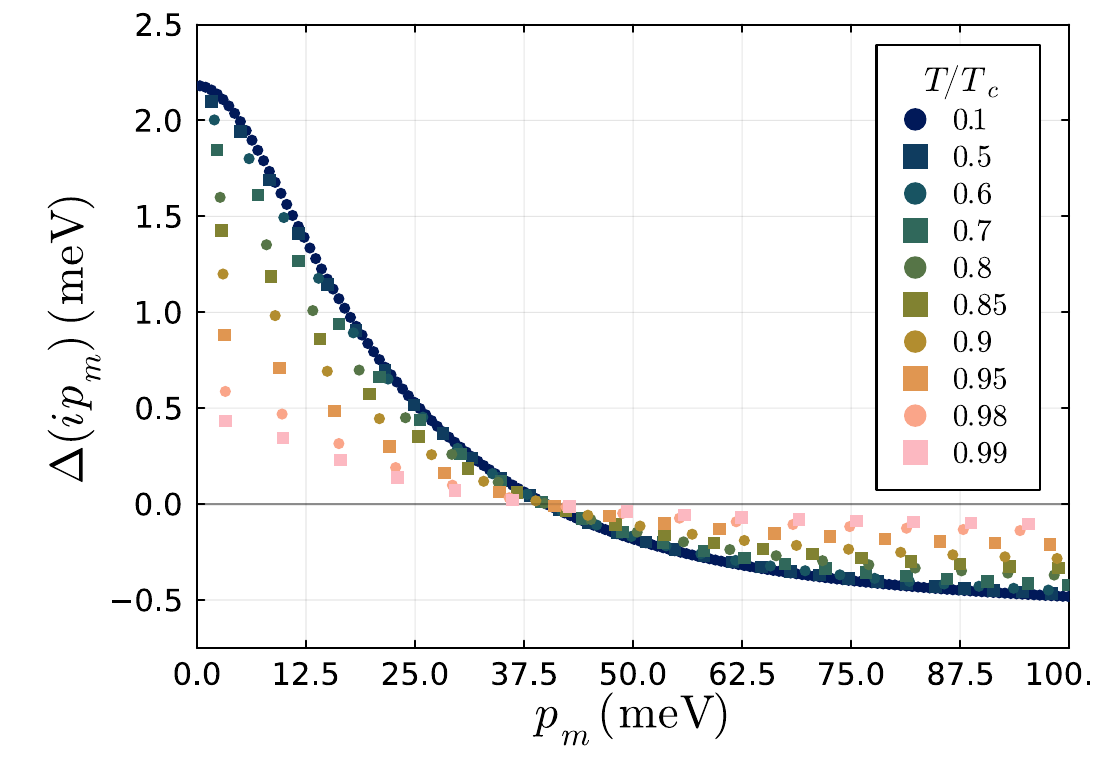}
    \caption{\textbf{The upper figure} displays the gap function evaluated on the imaginary Matsubara frequencies $p_m=(2m+1)\pi k_B T$ for a range of different values of $\lambda_2=\lambda_2(T=0)$, with $\lambda_1=1$, $T = 3$ K, $\LO=10$ meV, $\delta=1$ meV, $\nu_c=8$ meV, $u^\ast=0.12$ and $\omega_c=280$ meV. The nonlinear coupling $\lambda_2$ affects the normalization of the gap function, making it larger. Notice that the gap functions have a downward shift in both figures, making them pass through zero and approach a negative horizontal asymptote. This is solely due to the Coulomb repulsion. If $u^\ast=0$ the horizontal asymptote would be at zero (\ref{eq:asymptote_bandgap}). Practically the asymptote is `reached' around $p_m\approx 200$ meV, validating the choice of the cutoff $\omega_c=280$ meV.\\
    \noindent\textbf{The lower figure} shows the gap function for different temperatures for a fixed value of $\lambda_2(T=0)=0.2$, the other parameters are unchanged. The critical temperature is $T_c=12.26$ K for these parameter values. The spacing between the Matsubara frequencies is linear in the temperature. Naturally, a lower temperature corresponds to a higher absolute value of the gap function over the entire frequency space.}
    \label{fig:im-bandgap}
\end{figure}

Once the critical temperature of the superconductor is known, the superconducting gap function can be studied. The standard Eliashberg equations have to be solved first, yielding the normalization function $Z$ (\ref{eq:Z_isotropic}) and self-energy $W$ (\ref{eq:W_isotropic}) on the imaginary axis for a discrete set of points (the Matsubara frequencies). The gap function can be defined as $\Delta(ip_m)=W(ip_m)/Z(ip_m)$, which can be seen by looking at the poles of the Green's functions. Imaginary axis solutions for the gap function can be found in Figure \ref{fig:im-bandgap}.

The horizontal asymptote of the imaginary gap function can be calculated by setting $p_m\rightarrow+\infty$ in (\ref{eq:Z_isotropic}) and (\ref{eq:W_isotropic}). In this case $Z(ip_m)\rightarrow 1$ and $\lambda\rightarrow0$ (\ref{eq:lambda_p,k_Eliashberg}), so that\begin{align} \label{eq:asymptote_bandgap}
    \Delta(i\infty)=W(i\infty)=-\frac{u^\ast\pi}{\beta}\matssum{n}\frac{\Delta(ik_n)\theta(\omega_c-|k_n|)}{\sqrt{k_n^2+\Delta^2(ik_n)}}.
\end{align}
It goes to zero in absence of the Coulomb repulsion. It also has a different sign than the summation, or equivalently, the low frequency gap function. This is because of the summand receiving the biggest contributions for low frequencies $k_n$ due to the denominator. Addition of the nonlinear coupling increases the normalization of the gap function which shifts the horizontal asymptote more downward. Apart from this the gap function remains unchanged to the harmonic case.

\subsection{The real axis gap and normalization function}
\begin{figure*}
    \centering
    \includegraphics[scale=0.45]{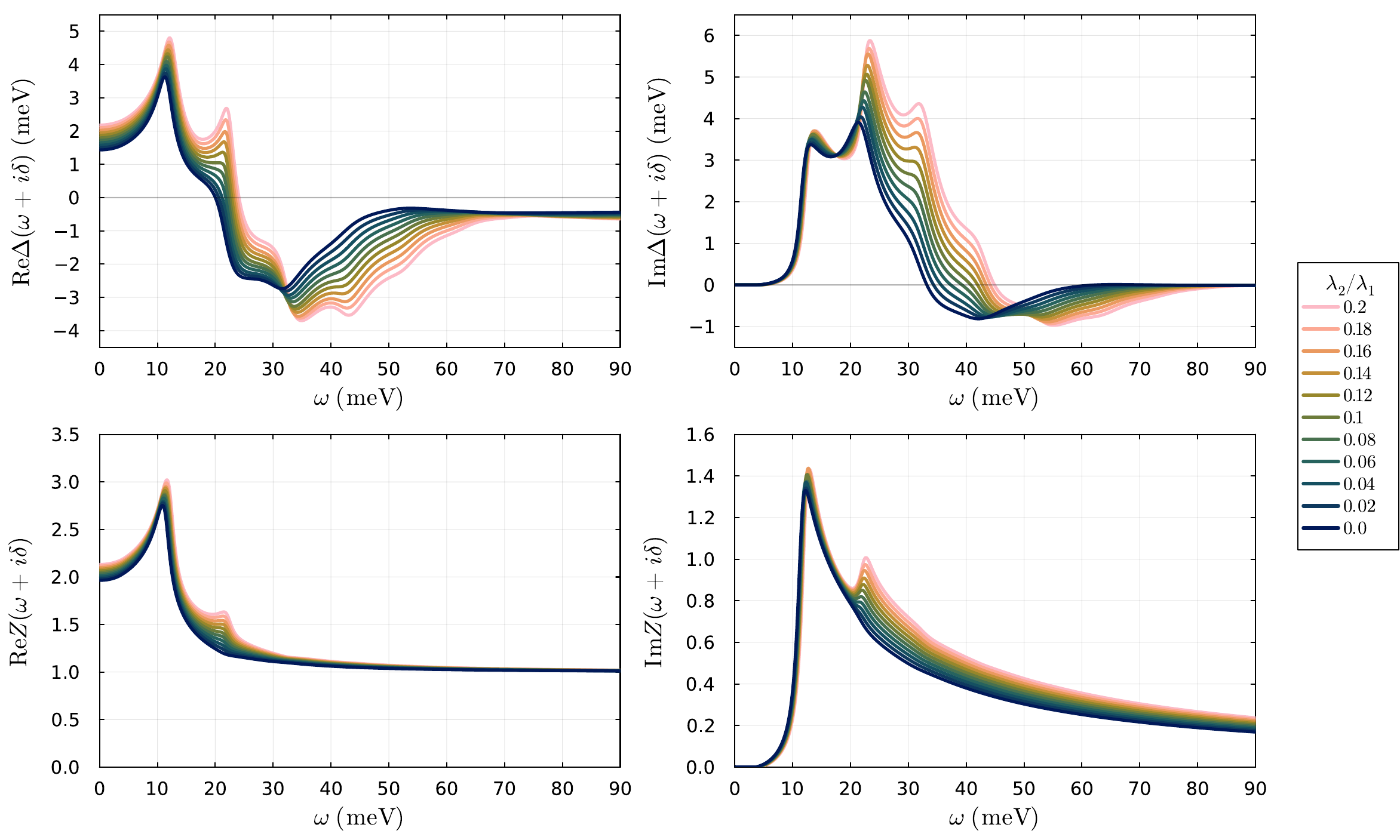}
    \includegraphics[scale=0.45]{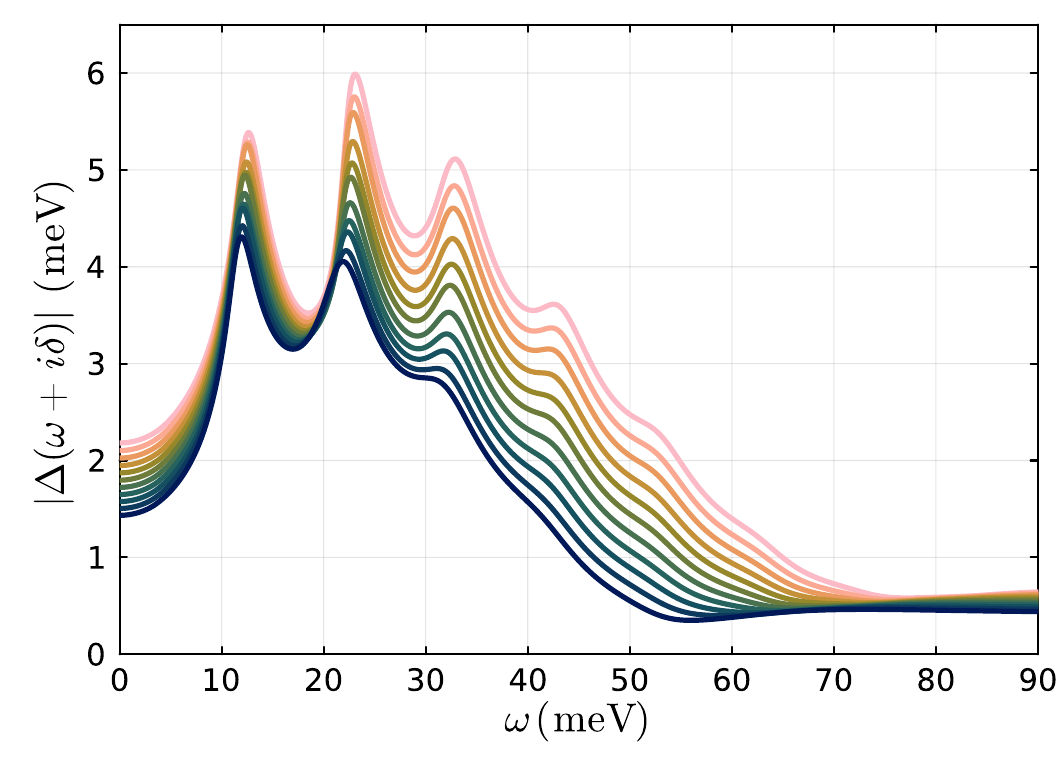}
    \caption{The real and imaginary parts of the superconducting gap function $\Delta(\omega+i\delta)$ and the normalization function $Z(\omega+i\delta)$ evaluated for real frequencies with $\lambda_1=1$, $T=3$ K, $\LO=10$ meV, $\delta=1$ meV, $\nu_c=8$ meV, $u^\ast=0.12$ and $\omega_c=280$ meV for different strengths of the nonlinear electron-phonon coupling $\lambda_2=\lambda_2(T=0)$. The lower figure displays the modulus of the gap function. Discussion of this figure is found in the text.}
    \label{fig:real-bandgap_lambda}
\end{figure*}

Having computed the solutions of the standard Eliashberg equations, the real-axis equations can be solved. On the real axis the gap function is defined as $\Delta(\omega+i\delta)=W(\omega+i\delta)/Z(\omega+i\delta)$, analogous to the definition on the imaginary axis. Unlike the imaginary axis formalism, the self-energies are now complex-valued. Figure \ref{fig:real-bandgap_lambda} shows the real and imaginary parts of the gap and normalization function.

The real-axis Eliashberg equations (\ref{eq:real-axis_Z}) and (\ref{eq:real-axis_W})  depend on the Eliashberg spectral function in two different ways: through the dimensionless coupling $\lambda$, but also explicitly in the integrations over the real axis. It is then no surprise that the gap and normalization functions display characteristics of the spectral function itself. The normalization function $Z$ adopts the two peaks at $\LO=10$ meV and $2\LO=20$ meV from the spectral function (Figure \ref{fig:Eliashberg_function}) directly with a minor shift towards higher frequencies. The gap function is more interesting. In the harmonic case $\lambda_2=0$ there is a distinct peak around the system's optical phonon frequency $\LO$, also with a minor shift. This peak is diminishingly repeated at every integer multiple of this frequency, taking into account the constant shift. In the nonlinear model this is also true. However, the peak at $2\LO$ receives contribution of the nonlinear peak in the Eliashberg spectral function as well, enlarging it notably. Then, every other peak at a multiple of $\LO$ is also affected by both peaks in the Eliashberg spectral function. Nonlinearity not only increases the gap function for all frequencies, it also extends the domain on which the gap function takes on a significant value, reaching its horizontal asymptote at a higher frequency. The higher the nonlinear coupling strength, the higher the contribution of these additional effects are. 
Through this self-consistent treatment of the self-energies it appears that the gap function has a maximum value of a few meV, never exceeding the $\LO$ energy of 10 meV. This, in combination with the normalization function being of order 1, at least consistently supports the assumption that the scale of the self-energies is that of the phonons.

\subsection{The gap function fix point}
The last quantity we will study is the fix point of the real part of the zero-temperature gap function, defined by\begin{align}
    \Delta_0=\text{Re}\Delta(\omega=\Delta_0),
\end{align}
which is simply the point where the real part of the gap function intersects the $\text{Re}\Delta(\omega)=\omega$ curve, as seen in Figure \ref{fig:real-bandgap_lambda}. Both BCS theory and conventional Migdal-Eliashberg theory predict that the fraction $2\Delta_0/(k_BT_c)$ approaches a constant value of 3.53 \cite{BCS, leavens1971ratio, leavens1973ratio} in the $T_c\rightarrow 0$ limit. An increase of the critical temperature modifies this value, with the general trend being described well by \cite{mitrovic1984ratio}\begin{align} \label{eq:fraction}
    \frac{2\Delta_0}{k_BT_c}=3.53\left[1+12.5\left(\frac{T_c}{\omega_{\ln}}\right)^2\ln\left(\frac{\omega_{\ln}}{2T_c}\right)\right],
\end{align}
where\begin{align}
    \omega_{\ln}=\exp\left[\frac{2}{\lambda_1+\lambda_2}\int d\nu \ln(\nu) \frac{\alpha^2F(\nu)}{\nu}\right]
\end{align}
is an average of the phonon energies in the superconductor. To study the ratio (\ref{eq:fraction}) between the gap and critical temperature using this model, we prefer to keep the general structure of the Eliashberg spectral function unchanged, without altering the phonon spectrum. Therefore, the only way to vary $T_c$ is to vary the electron-phonon coupling strength $\lambda=\lambda_1+\lambda_2$. It is important to keep in mind that the nonlinear electron-phonon coupling is temperature dependent through the Bose-Einstein distribution (\ref{eq:lambda2(T)}). This also means that $\omega_{\ln}$ will be dependent on the temperature.

\begin{figure}
    \centering
    \includegraphics[scale=0.375]{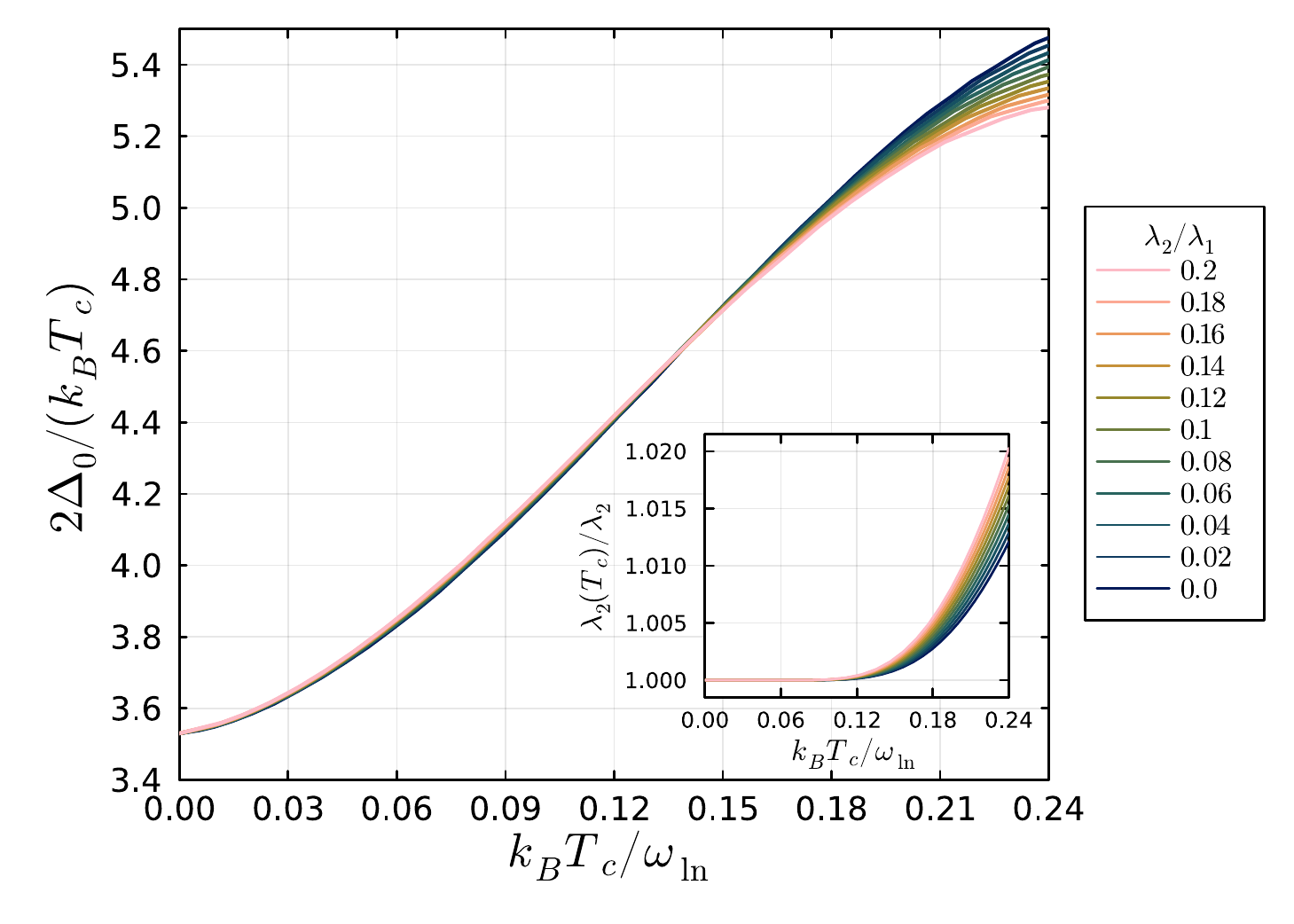}
    \caption{\textbf{The main figure} shows the fix point of the zero-temperature gap function divided by the critical temperature for $\lambda_1\in[\:\!0,\:3.2\:\!]$ (corresponding to $T_c\in[\:\!0\:\text{K},\: 26.2$ K$\:\!]$ for $\lambda_2(0)/\lambda_1 = 0$) and various $\lambda_2(0)$ values, the rest of the parameters are unchanged compared to previous figures.\\
    \noindent\textbf{The inset} displays the relative temperature scaling of the nonlinear electron-phonon coupling. Of course the critical temperature rises with increasing nonlinear coupling. Therefore, a higher nonlinear coupling is affected more heavily by the temperature scaling.} 
    \label{fig:fraction}
\end{figure}

Figure \ref{fig:fraction} shows the results for $2\Delta_0/(k_BT_c)$ in the nonlinear model. We see that the nonlinear coupling does not affect the relation between these quantities in the low temperature regime ($k_BT_c/\omega_{\ln}\approx0.15$ corresponds with a critical temperature of around $T_c\approx17\:\text{K}\sim19$ K in our model). For higher critical temperatures, the temperature scaling of $\lambda_2(T)$ becomes important and the curves start to diverge. This is to be expected, because the fix point of the zero-temperature band gap $\Delta_0$ only takes into account $\lambda_2(0)$, while the critical temperature feels the effect of an enhanced $\lambda_2(T_c)$. This discrepancy will only become larger in the higher temperature regime. Notice also that the averaged phonon frequency $\omega_{\ln}$ in the spectrum is higher for larger $\lambda_2$ values, since the weight is shifted more towards higher frequencies. This means that the same value of $k_BT_c/\omega_{\ln}$ corresponds to a higher critical temperature for stronger nonlinear coupling.

\section{Summary and outlook}
In this paper we constructed the Eliashberg equations using a Hamiltonian which includes the electron-phonon coupling up to second order (\ref{eq:Hamiltonian}). We derived a correction term to the harmonic Eliashberg spectral function. Addition of nonlinearity of this form to Migdal-Eliashberg theory does not affect the standard or real-axis description of the Eliashberg equations. Only the Eliashberg spectral function $\alpha^2F_\mathbf{p,k}(\nu)$ is affected by nonlinearity, introducing an additional term solely describing the nonlinear part (\ref{eq:Eliashberg_function}). The nonlinear part becomes more dominant for higher temperatures, so superconductors with high superconducting critical temperatures might benefit from this description even more. This nonlinear and anisotropic expression for the spectral function in combination with the unaltered Eliashberg equations can be seen as the main result of this manuscript.

It is a general result which holds for any material where the phonon energy scale and the Debye-Waller self-energy are negligible compared to the maximum of the electronic band, and where the Migdal approximation is still valid. Additionally, we assume a quasi constant density of states near the Fermi surface and only one relevant electronic conduction band. If the 1-electron-2-phonon matrix element is known and the superconducting properties of this nonlinear material are desired, it is important to check if the Debye-Waller energy is indeed a small quantity, at least compared to the electronic conduction band. If this is not the case, then this model has to be modified to include the energy shift $\chi(\mathbf{p})=\Sigma^\text{DW}(\mathbf{p})$.

Ab initio calculations of the phonon spectra and electron-phonon matrix elements can be difficult with inclusion of nonlinearity. While anharmonic phonon-phonon interactions are already widely implemented, using SSCHA methods \cite{Monacelli_2021} for example, nonlinear electron-phonon coupling is often overlooked. There are ways to include this as well \cite{bianco2023}, but a direct link with known theory or microscopic quantities is lacking, limiting predictive power or fundamental understanding. Use of Monte Carlo models which simulate anharmonic phonon-phonon interactions and nonlinear electron-phonon coupling have already been shown to improve descriptions in polaron physics, accurately explaining the temperature dependence of polaron mobility in the lead halide perovskite MAPbI$_3$ for example \cite{Saidi2016}.

Recently, a semi-analytical expression to calculate the long-range part of the 1-electron-2-phonon matrix element has been derived \cite{houtput2024}. This is another step towards making nonlinear ab initio calculations more tractable, starting from a microscopic theory. Therefore, an explicit expression of the input needed to determine the superconducting properties of a material which displays signs of significant anharmonicity in Eliashberg theory can be of great use. 

We reiterate that this model does not include phonon-phonon interactions, or so-called phonon anharmonicity. This can be added by including a phonon-phonon interaction term in the Hamiltonian. Inclusion of the latter does not change the theory in any fundamental manner, as it simply renormalizes the phonon energies and modifies the isotope effect. It also allows some additional diagrams for which the propagators can be corrected, shown in Figure \ref{fig:diagram3ph}. These diagrams might provide a significant contribution as well, but were not considered in this manuscript. The toy model Eliashberg function used in Sec. \ref{sec:Numerics} to illustrate the consequence of nonlinear electron-phonon interaction is easily numerically tractable, but it is most likely oversimplified and will not yield any qualitative description of real materials. It does outline the core effects of the nonlinear electron-phonon coupling, such as an increase of the gap function evaluated on the imaginary axis and modifications of the gap and normalization function on the real axis. The temperature dependence of the nonlinear electron-phonon coupling $\lambda_2(T)$ shows that the ratio $\frac{2\Delta_0}{k_BT_c}$ is affected at higher critical temperatures, where a larger nonlinear coupling $\lambda_2$ causes a larger disparity from conventional results. Most notably, we see a monotonic increase in the superconducting critical temperature $T_c$ as a function of the 1-electron-2-phonon coupling strength, which can become even more significant for superconductors with higher critical temperatures due to the temperature scaling of the nonlinear coupling.

\section*{Acknowledgments}
This work is supported financially by the Research Foundation–Flanders (FWO), projects numbers G0A9F25N, G0AIY25N, GOH1122N,  G061820N, and G060820N, and by the University Research Fund (BOF) of the University of Antwerp. I. Z. acknowledges support from the Research Foundation-Flanders (FWO) through a Ph.D. fellowship in fundamental research, project number 1120825N. M. H. acknowledges support from the Research Foundation-Flanders (FWO) through a postdoctoral fellowship in fundamental research, project number 1224724N.

\appendix
\section{Calculating the Dyson series} \label{app:Dyson}
This first part of the Appendix will be used to go over the method of constructing the Dyson series used as starting point for setting up the Eliashberg equations. This will be done by explicitly calculating the Dyson series for the novel term in the Hamiltonian (\ref{eq:Hamiltonian}), being the 1-electron-2-phonon interaction. The method of calculating the Coulomb or 1-electron-1-phonon Dyson series is completely analogous. A first order $S$-matrix expansion can be performed to receive the Debye-Waller diagram, shown in Figure \ref{fig:debye-waller}. However, its contribution can be shown to be vanishing (Appendix \ref{app:Debye-Waller}). Therefore, we start with a second order expansion on the propagators\begin{widetext}\begin{align}
    \Gn{2}{p}{\tau}=-\frac{(-1)^2}{2!}\intbeta d\tau_1\intbeta d\tau_2\Tbrak{\anii{p}{\ua}(\tau)\hat{V}(\tau_1)\hat{V}(\tau_2)\cree{p}{\ua}(0)}\label{eq:G_2d_order}\\
    \Fdn{2}{p}{\tau}=\frac{(-1)^2}{2!}\intbeta d\tau_1\intbeta d\tau_2\Tbrak{\cree{p}{\ua}(\tau)\hat{V}(\tau_1)\hat{V}(\tau_2)\cree{-p}{\da}(0)}
\end{align}\end{widetext}
Now, the interactions of which the interacting part of the Hamiltonian is composed (\ref{eq:H_int}) can be inserted, allowing for contributions of the 1-electron-1-phonon coupling $\hat{V}_g$, the 1-electron-2-phonon coupling $\hat{V}_\gamma$ and the Coulomb interaction $\hat{V}_u$. Overall, the expectation value consists of many contributions\begin{align}
\hat{V}^2=\hat{V}_g^2+\hat{V}_\gamma^2+\hat{V}_u^2+\ldots.
\end{align}
All mixing terms vanish, since they leave us with either an expectation value over an odd amount of phonon propagators, or a Debye-Waller diagram mixed with a Coulomb interaction. The quadratic term of the Coulomb interaction is already of higher order, because the Coulomb interaction is assumed to be instantaneous. This means that a first order expansion already allowed us to correct the propagators for the Coulomb interaction. This correction is then promoted to a Dyson series, which repeatedly includes the interaction up to infinite order. The only difference that a quadratic term can make is to allow us to include vertex corrections, which we neglect following the Migdal approximation \cite{Migdal}. The only contributing factors to the second order expansion of the propagators are those with electron-phonon coupling. We continue with calculating the Dyson series for the nonlinear coupling explicitly, since this is the essential term of this manuscript. Filling in the expression of $\hat{V}_\gamma$ in (\ref{eq:G_2d_order}) yields\begin{widetext}\begin{align}
    \Gn{2\textrm{A}}{p}{\tau}=-\frac{1}{8\nu^2}\sum_{\mathbf{k},\mathbf{k'},\mathbf{q}}\sum_{\mathbf{b},\mathbf{b'},\mathbf{q'}}&\sum_{\sigma,\sigma'}\intbeta d\tau_1 \intbeta d\tau_2 \gamma(\mathbf{k,k',-q})\gamma(\mathbf{b,b',-q'}) \Tbrak{\A{k'}(\tau_1)\A{-q}(\tau_1)\A{b'}(\tau_2)\A{-q'}(\tau_2)}\nonumber \\
    &\times\Tbrak{\anii{p}{\ua}(\tau)\cree{k+k'-q}{\sigma}(\tau_1)\anii{k}{\sigma}(\tau_1)\cree{b+b'-q'}{\sigma'}(\tau_2)\anii{b}{\sigma'}(\tau_2)\cree{p}{\ua}(0)}.
\end{align}
For the phonon part of the expectation value we only consider pairing of operators at different times to avoid Debye-Waller diagrams. The expectation value can be decomposed following Wick's theorem, since we do not include phonon-phonon interactions. The decomposition gives the same contribution twice, $\D{k'}{\tau_1-\tau_2}\D{-q}{\tau_1-\tau_2}$, since $\gamma^\ast(\mathbf{b,k',-q})=\gamma(\mathbf{b,-k',q)}=\gamma(\mathbf{b,-q,k})$, which is easily derived from the Hamiltonian being hermitian. After a Wick decomposition on the electronic expectation value and using that $\F{p}{\tau}=\Fd{p}{\tau}=\Fd{-p}{\tau}=\Fd{p}{-\tau}$ \cite{Mahan}, one obtains\begin{align}
    \Gn{2\textrm{A}}{p}{\tau}=-\frac{1}{4\nu^2}\sum_{\mathbf{k'},\mathbf{q}}\intbeta d\tau_1 \intbeta& d\tau_2 \gamma(\mathbf{p,k',q})\gamma^\ast(\mathbf{p,k',q})\D{k'}{\tau_1-\tau_2}\D{q}{\tau_1-\tau_2}\nonumber \\
    &\times2\begin{bmatrix*}[l]
        -\Gn{0}{p}{\tau_2}\Gn{0}{p-k'-q}{\tau_1-\tau_2}\Gn{0}{p}{\tau-\tau_1}\vspace{0.2cm}\\
        +\Fdn{0}{p}{\tau_2}\Fdn{0}{p-k'-q}{\tau_1-\tau_2}\Gn{0}{p}{\tau-\tau_1}\vspace{0.2cm}\\
        +\Gn{0}{p}{\tau_2}\Fdn{0}{p-k'-q}{\tau_1-\tau_2}\Fdn{0}{p}{\tau-\tau_1}\vspace{0.2cm}\\
        +\Fdn{0}{p}{\tau_2}\Gn{0}{-p+k'+q}{\tau_2-\tau_1}\Fdn{0}{p}{\tau-\tau_1}
        \end{bmatrix*}_{\textstyle \raisebox{3pt}{.}}
\end{align}
The electronic part gives four different contributions, with each of them being generated twice from the Wick expansion. This follows from the fact that there is inherently no difference between the two interaction terms, being equal but acting at different times. This is also the reason that now $\mathbf{p}=\mathbf{b}=\mathbf{k}$. As seen in this expression it is crucial to include Wick pairings over the anomalous propagators. One can now change the momentum summation over $\mathbf{k'}$ to one over $\mathbf{k}=\mathbf{p-k'-q}$. The whole can also be transformed to Fourier space, yielding\begin{align}
    \Gn{2\textrm{A}}{p}{ip_m}=\frac{1}{2\nu^2\beta^2}\sum_{\mathbf{k},\mathbf{q}}\matssum{n,l}&\big|\gamma(\mathbf{p,p-k-q,q})\big|^2\D{p-k-q}{ip_m-ik_n-iq_l}\D{q}{iq_l}\nonumber\\
    &\times\begin{bmatrix*}[l]
        +\Gn{0}{p}{ip_m}\Gn{0}{k}{ik_n}\Gn{0}{p}{ip_m}\vspace{0.2cm}\\
        -\Fdn{0}{p}{ip_m}\Fdn{0}{k}{ik_n}\Gn{0}{p}{ip_m}\vspace{0.2cm}\\
        -\Gn{0}{p}{ip_m}\Fdn{0}{k}{ik_n}\Fdn{0}{p}{ip_m}\vspace{0.2cm}\\
        -\Fdn{0}{p}{ip_m}\Gn{0}{-k}{-ik_n}\Fdn{0}{p}{ip_m}
        \end{bmatrix*}_{\textstyle \raisebox{3pt}{.}}
\end{align}
The expression above can be simplified heavily by noticing common terms, if we define\begin{align}
    \Sigma^{(0)}_\gamma(\mathbf{p},ip_m)=\frac{1}{2\beta^2}\matssum{n,l}\int\frac{d\mathbf{k}}{(2\pi)^3}\int\frac{d\mathbf{q}}{(2\pi)^3}\big|\gamma(\mathbf{p,p-k-q,q})\big|^2\D{p-k-q}{ip_m-ik_n-iq_l}\D{q}{iq_l}\Gn{0}{k}{ik_n},\\
    \phi^{(0)}_\gamma(\mathbf{p},ip_m)=\frac{1}{2\beta^2}\matssum{n,l}\int\frac{d\mathbf{k}}{(2\pi)^3}\int\frac{d\mathbf{q}}{(2\pi)^3}\big|\gamma(\mathbf{p,p-k-q,q})\big|^2\D{p-k-q}{ip_m-ik_n-iq_l}\D{q}{iq_l}\Fdn{0}{k}{ik_n},
\end{align}\end{widetext}
this becomes\begin{align}
    \mathcal{G}^{(2\textrm{A})}(p)&=\mathcal{G}^{(0)}(p)\left[\Sigma_\gamma^{(0)}(p)\mathcal{G}^{(0)}(p)-\phi_\gamma^{(0)}(p)\mathcal{F}^{(0)\dagger}(p)\right]\nonumber\\
    &-\mathcal{F}^{(0)\dagger}(p)\left[\phi_\gamma^{(0)}(p)\mathcal{G}^{(0)}(p)-\Sigma_\gamma^{(0)}(-p)\mathcal{F}^{(0)\dagger}(p)\right].
\end{align}
Here, a shorter notation was used where $p=(\mathbf{p},ip_m)$. We changed the momentum summation to integrations by assuming $\nu$ to be large, and we also used that $\phi_\gamma=\phi_\gamma^\dagger$, this originates from the anomalous propagators being equal. Now that the lowest order contribution of the nonlinear electron-phonon coupling term has been calculated, the expression can be promoted to a Dyson series. For the electron propagator this yields\begin{align}
    \mathcal{G}(p)&=\mathcal{G}^{(0)}(p)\left[1+\Sigma_\gamma^{(0)}(p)\mathcal{G}(p)-\phi_\gamma^{(0)}(p)\mathcal{F}^{\dagger}(p)\right]\nonumber\\
    &-\mathcal{F}^{(0)\dagger}(p)\left[\phi_\gamma^{(0)}(p)\mathcal{G}(p)+\Sigma_\gamma^{(0)}(-p)\mathcal{F}^{\dagger}(p)\right].
\end{align}
Similarly, the Dyson series for the anomalous propagator can be found\begin{align}
    \mathcal{F}^\dagger(p)&=\mathcal{F}^{(0)\dagger}(p)\left[1+\Sigma_\gamma^{(0)}(p)\mathcal{G}(p)-\phi_\gamma^{(0)}(p)\mathcal{F}^\dagger(p)\right]\nonumber\\
    &+\mathcal{G}^{(0)}(-p)\left[\Sigma_\gamma^{(0)}(-p)\mathcal{F}^\dagger(p)+\phi_\gamma^{(0)}(p)\mathcal{G}(p)\right]
\end{align}
Now, one has to realise that the anomalous propagators are only to be assumed non-zero in the superconducting state, for a normal metal they are omitted. The superconductivity that we are describing is a result of the electron pairs being bound together through phonon coupling. A zeroth order propagator is a free propagator, unaffected by any interactions which might be described by the Hamiltonian. For this reason, the zeroth order anomalous propagators have to vanish. However, one problem that arises is that the anomalous self-energy will be omitted as well, all the while it is clear from the propagator's Dyson series that it does add contributions. For this reason, the self-energies have to be redefined to be dependent of the corrected propagators instead of the uncorrected ones. Doing this results in the following two Dyson series\begin{align}
    \mathcal{G}(p)=\mathcal{G}^{(0)}(p)\left[1+\Sigma_\gamma(p)\mathcal{G}(p)-\phi_\gamma(p)\mathcal{F}^\dagger(p)\right],\\
    \mathcal{F}^\dagger(p)=\mathcal{G}^{(0)}(-p)\left[\Sigma_\gamma(-p)\mathcal{F}^\dagger(p)+\phi_\gamma(p)\mathcal{G}(p)\right],
\end{align}
where the self-energies are now defined as (\ref{eq:Sigma_gamma}) and (\ref{eq:Omega_gamma}). Take note that the anomalous propagators are set equal. These Dyson series are of the same form as for the Coulomb interaction and the 1-electron-1-phonon interaction. This is easily seen by realising that the only difference with this derivation and an harmonic one is the fact that there is only one phonon operator in the latter case, so that the self-energies only obtain contributions of a single phonon propagator without affecting the electronic part.

\section{Anisotropic Eliashberg equations} \label{app:Anisotropic}
We will go over the derivation of the anisotropic Eliashberg equations, since it is important to be aware of all approximations that have to be made and where they come from. This derivation reproduces the work of \cite{Mahan, Allen, Margine_Giustino}, which is once again possible because only the effective potential changes in our model, which is kept general in the derivation. From the Dyson series (\ref{eq:Dyson_G}) and (\ref{eq:Dyson_F}) one can start constructing the Eliashberg equations. First, the coupled equations can be solved to obtain\begin{align}
    \mathcal{G}(p)=\frac{ip_mZ(p)+\chi(p)+\varepsilon_\mathbf{p}}{\left[ip_mZ(p)\right]^2-\left[\chi(p)+\varepsilon_\mathbf{p}\right]^2-W^2(p)},\label{eq:G_self-consistent}\\
    \mathcal{F}^\dagger(p)=\frac{-W(p)}{\left[ip_mZ(p)\right]^2-\left[\chi(p)+\varepsilon_\mathbf{p}\right]^2-W^2(p)},\label{eq:F_self-consistent}
\end{align}
where the self-energy $S(p)$ has been decomposed into a symmetric and an antisymmetric part in $ip_m$\begin{align} \label{eq:S(p)_decomposition}
    S(p)=ip_m\left[1-Z(p)\right]+\chi(p).
\end{align}
Notice that even though $Z(p)$ is a component of the antisymmetric part of $S(p)$, it is symmetric due to factoring out an additional $ip_m$. $\chi(p)$ is defined as the symmetric component. The effective potential is symmetric in $ip_m$. The anomalous self-energy $W(k)$ is symmetric in its frequency argument as well. This comes from the fact that the anomalous propagators are set equal, which made them independent of the sign of their time (thus frequency after Fourier transformation) argument. The expressions for the propagators can be substituted in the definitions of the self-energies (\ref{eq:S(p)_V_eff}) and (\ref{eq:W(p)_V_eff}), yielding three self-consistent equations\begin{gather}
    ip_m\left[1-Z(p)\right]=\frac{1}{\beta}\matssum{n}\int\frac{d\mathbf{k}}{(2\pi)^3}V_\text{eff}(p,k)\frac{ik_nZ(k)}{\Xi(k,\varepsilon_\mathbf{k})},\label{eq:S(p)_antisymm}\\
    \chi(p)=\frac{1}{\beta}\matssum{n}\int\frac{d\mathbf{k}}{(2\pi)^3}V_\text{eff}(p,k)\frac{\chi(k)+\varepsilon_\mathbf{k}}{\Xi(k,\varepsilon_\mathbf{k})},\label{eq:S(p)_symm}\\
    W(p)=-\frac{1}{\beta}\matssum{n}\int\frac{d\mathbf{k}}{(2\pi)^3}V_\text{eff}(p,k)\frac{W(k)}{\Xi(k,\varepsilon_\mathbf{k})},
\end{gather}
where 
\begin{align}
    \Xi(\mathbf{k},ik_n,\varepsilon_\mathbf{k})=k_n^2Z^2(k)+\left[\chi(k)+\varepsilon_\mathbf{k}\right]^2+W^2(k).
\end{align}
The equations for $Z(p)$ and $\chi(p)$ both come from the equation of $S(p)$. By splitting the right-hand and left-hand side of the self-consistent equation into a symmetric and an antisymmetric part, they can be split into two separate equations. The procedure is to treat $\chi$ first, which can be rewritten as\begin{align}
    \chi(p)=\frac{1}{\beta}\matssum{n}\int\frac{d\mathbf{k}}{(2\pi)^3}\int_{-\epsilon}^{+\epsilon}dEV_\text{eff}(p,k)\frac{\chi(k)+E}{\Xi(k,E)}\gamma_\mathbf{k}(E),
\end{align}
with $\gamma_\mathbf{k}(E)=\delta(E-\varepsilon_\mathbf{k})$. Let us assume that the self-energies $Z$, $\chi$ and $W$ vary slowly in function of the electronic band energy $\varepsilon_\mathbf{k}=E$, at least in comparison with the explicit presence of $E$ in the fraction. A physical interpretation for this approximation is the fact that the phonon energies act on a small energy window around the Fermi energy $[\epsilon_F-\omega_D;\epsilon_F+\omega_D]$, meaning that the self-energies can simply be evaluated at the Fermi energy. Then, one can set $\delta(E-\varepsilon_\mathbf{k})\rightarrow\delta(\epsilon_F-\varepsilon_\mathbf{k})=\gamma_\mathbf{k}$ \cite{Allen} and perform the energy integral analytically\begin{align}
    \chi(p)=\frac{1}{\beta}\matssum{n}\int\frac{d\mathbf{k}}{(2\pi)^3}\int_{-\epsilon+\chi(k)}^{+\epsilon+\chi(k)}d\mathcal{E}\frac{\gamma_\mathbf{k}V_\text{eff}(p,k)\mathcal{E}}{\Xi(k,\mathcal{E}-\chi(k))}.\label{eq:chi(p)_analytic-integral}
\end{align}
The integrand is antisymmetric in $\mathcal{E}=E+\chi(k)$, with symmetric boundaries the integral would equal zero. If we assume that the electronic band has a high maximum energy $\epsilon$, then it would make no difference if we integrate to this maximum or infinity due to the rapid convergence of the integrand (being on the energy scale of the phonons). Therefore, we will let $\epsilon\rightarrow+\infty$, making the boundaries more and more symmetric, letting $\chi(p)\rightarrow 0$ in a self-consistent manner. Note that this also assumes that $\chi(p)\ll\epsilon$ before making the extensions to infinity. Because $\chi(p)$ acts on the level of the phonon energy, this is already captured by previous approximations.

Only two self-consistent equations remain, where it is possible to use $\chi=0$ and introduce the integral over the energy. Using the same approximations as before one can perform the integrals analytically again, resulting in\begin{widetext}\begin{gather}
    Z(p)=1-\frac{2}{ip_m\beta}\matssum{n}\int\frac{d\mathbf{k}}{(2\pi)^3}\gamma_\mathbf{k}V_\text{eff}(p,k)\frac{ik_nZ(k)}{\sqrt{k_n^2Z^2(k)+W^2(k)}}\text{Arctan}\left[\frac{\epsilon}{\sqrt{k_n^2Z^2(k)+W^2(k)}}\right],\\
    W(p)=-\frac{2}{\beta}\matssum{n}\int\frac{d\mathbf{k}}{(2\pi)^3}\gamma_\mathbf{k}V_\text{eff}(p,k)\frac{W(k)}{\sqrt{k_n^2Z^2(k)+W^2(k)}}\text{Arctan}\left[\frac{\epsilon}{\sqrt{k_n^2Z^2(k)+W^2(k)}}\right].
\end{gather}
The limit of $\epsilon\rightarrow+\infty$ has not been taken yet, doing this would change the inverse tangent functions to constants, simplifying the coupled self-consistent equations in the process. However, the effective potential is composed of a part that describes Coulomb interactions and one that describes the electron-phonon coupling (\ref{eq:lambda_p,k}). In this expression, it is important to note that the Coulomb interaction $u(\mathbf{p-k})$ acts instantaneous in this model, making it independent of any Matsubara frequency. Therefore, if one extends the integration boundaries to infinity, `canceling' the inverse tangent, the Matsubara summation for the Coulomb term will diverge. The workaround is to set $\epsilon\rightarrow+\infty$ anyway and introduce a cutoff for this term, $\theta(\epsilon-|k_n|)$, since normally the inverse tangent acts a soft cutoff at $\epsilon$. So\begin{gather}
    Z(p)=1-\frac{\pi}{ip_m\beta}\matssum{n}\int\frac{d\mathbf{k}}{(2\pi)^3}\gamma_\mathbf{k}\left[u(\mathbf{p-k})\theta(\epsilon-|k_n|)+V_\text{ph}(\mathbf{p,k},ip_m-ik_n)\right]\frac{ik_nZ(k)}{\sqrt{k_n^2Z^2(k)+W^2(k)}},\\
    W(p)=-\frac{\pi}{\beta}\matssum{n}\int\frac{d\mathbf{k}}{(2\pi)^3}\gamma_\mathbf{k}\left[u(\mathbf{p-k})\theta(\epsilon-|k_n|)+V_\text{ph}(\mathbf{p,k},ip_m-ik_n)\right]\frac{W(k)}{\sqrt{k_n^2Z^2(k)+W^2(k)}}.
\end{gather}
The fraction in the equation for $Z(p)$ is antisymmetric in the Matsubara frequency $k_n$. The Coulomb term is symmetric in this frequency, making it vanish when the summation is performed. In the equation for $W$ the fraction is symmetric, so this argument can not be used. If one now defines $V_\text{ph}(\mathbf{p,k},ip_m-ik_n)=-\lambda_\mathbf{p,k}(ip_m-ik_n)$ and assumes the Coulomb force to act isotropically (or take an isotropic averaging (\ref{eq:u_isotropic})) $u(\mathbf{p-k})\rightarrow u$, then the anisotropic Eliashberg equations are obtained\begin{gather}
    Z(p)=1+\frac{\pi}{ip_m\beta}\matssum{n}\int\frac{d\mathbf{k}}{(2\pi)^3}\frac{i\gamma_\mathbf{k}k_nZ(k)}{\sqrt{k_n^2Z^2(k)+W^2(k)}}\lambda_\mathbf{p,k}(ip_m-ik_n),\\
    W(p)=\frac{\pi}{\beta}\matssum{n}\int\frac{d\mathbf{k}}{(2\pi)^3}\frac{\gamma_\mathbf{k}W(k)}{\sqrt{k_n^2Z^2(k)+W^2(k)}}\left[\lambda_\mathbf{p,k}(ip_m-ik_n)-u\theta(\epsilon-|k_n|)\right].
\end{gather}\end{widetext}
\section{Contributions of the Debye-Waller diagram} \label{app:Debye-Waller}
The first order $S$-matrix expansion resulted in a contribution of the Coulomb interaction, but not for the phonons. For the 1-electron-1-phonon interaction this is exact, since the expectation value over a single phonon operator $\A{q}$ vanishes. For the 1-electron-2-phonon interaction, one can actually calculate a contribution. This contribution is one for the Debye-Waller diagram, as seen in Figure \ref{fig:debye-waller}. In this Appendix we will study the effects of this diagram on the Eliashberg equations, starting with the first order expansion of the electron propagator\begin{align}
    &\Gn{\text{DW}}{p}{\tau}=-\frac{1}{2\nu}\sum_\mathbf{k,q}\sum_\sigma \gamma(\mathbf{k,q,q})\D{q}{0}\nonumber\\
    &\times\intbeta d\tau_1 \Tbrak{\anii{p}{\ua}(\tau)\cree{k}{\sigma}(\tau_1)\anii{k}{\sigma}(\tau_1)\cree{p}{\ua}(0)}.
\end{align}
Even though the diagram has a 1-electron-2-phonon interaction vertex, the `two' phonons are actually the same phonon, which annihilates itself instantaneously. Performing a Wick expansion on the electronic expectation value gives\begin{align}
    &\Gn{\text{DW}}{p}{\tau}=-\frac{1}{2\nu}\sum_\mathbf{q}\gamma(\mathbf{p,q,q})\D{q}{0}\intbeta d\tau_1\nonumber\\
    &\times \left[ \Gn{0}{p}{\tau-\tau_1}\Gn{0}{p}{\tau_1}-\Fn{0}{p}{\tau-\tau_1}\Fdn{0}{p}{\tau_1} \right],
\end{align}
which can be transformed into Fourier space\begin{align}
    &\Gn{\text{DW}}{p}{ip_m}=-\frac{1}{2\nu\beta}\matssum{n}\sum_\mathbf{q}\gamma(\mathbf{p,q,q})\D{q}{iq_n}\nonumber\\
    &\times \left[ \Gn{0}{p}{ip_m}\Gn{0}{p}{ip_m}-\Fn{0}{p}{ip_m}\Fdn{0}{p}{ip_m} \right].
\end{align}
Defining the Debye-Waller self-energy as the common term\begin{align}
    \Sigma^\text{DW}(\mathbf{p})&=-\frac{1}{2\beta}\matssum{n}\int\frac{d\mathbf{q}}{(2\pi)^3}\gamma(\mathbf{p,q,q})\D{q}{iq_n}\nonumber\\
    &=-\frac{1}{2}\int\frac{d\mathbf{q}}{(2\pi)^3}\gamma(\mathbf{p,q,q})(1+2n_B^\mathbf{q})
\end{align}
allows one to promote the correction to a Dyson series by adding the zeroth order contribution and inserting the `infinitely corrected' propagators into the left-hand and right-hand side, after which the zeroth order anomalous propagator can be set to zero,\begin{align}
    \mathcal{G}(p)=\mathcal{G}^{(0)}(p)\left[1+\Sigma^\text{DW}(\mathbf{p})\mathcal{G}(p)\right].
\end{align}
One can follow the same procedure to obtain the Dyson series for the anomalous propagator\begin{align}
    \mathcal{F}^\dagger(p)=\mathcal{G}^{(0)}(-p)\Sigma^\text{DW}(\mathbf{-p})\mathcal{F}^\dagger(p).
\end{align}
These Dyson equations are not of the usual structure for the superconducting system. The Debye-Waller diagram does not have an anomalous self-energy contribution, so $\phi^\text{DW}(\mathbf{p})=0$, which decouples the series. To include the Debye-Waller diagram, the self-energy can be added to the electron self-energy $S(p)$ (\ref{eq:S(p)}), yielding\begin{align}
    S(p)=\Sigma_u(p)+\Sigma_g(p)+\Sigma_\gamma(p)+\Sigma^\text{DW}(\mathbf{p}).
\end{align}
Normally, the self-energies are dependent on the electron propagator as well, which allowed us to rewrite them using an effective potential (\ref{eq:S(p)_V_eff}). The Debye-Waller term can not be written with an effective potential in this way, its contribution will remain an addition to the original self-energy. This contribution is added purely to the symmetric part of $S(p)$, which we called $\chi(p)$ (\ref{eq:S(p)_decomposition}). So, after filling in the electron propagator (\ref{eq:G_self-consistent}) and splitting the symmetric and antisymmetric parts of $S(p)$ into two separate equations, one reobtains (\ref{eq:S(p)_antisymm}) and (\ref{eq:S(p)_symm})\begin{gather}
    Z(p)=1-\frac{1}{ip_m\beta}\matssum{n}\int\frac{d\mathbf{k}}{(2\pi)^3}V_\text{eff}(p,k)\frac{ik_nZ(k)}{\Xi(k,\varepsilon_\mathbf{k})},\\
    \chi(p)=\frac{1}{\beta}\matssum{n}\int\frac{d\mathbf{k}}{(2\pi)^3}V_\text{eff}(p,k)\frac{\chi(k)+\varepsilon_\mathbf{k}}{\Xi(k,\varepsilon_\mathbf{k})}+\Sigma^\text{DW}(\mathbf{p}),
\end{gather}
with the Debye-Waller contribution now added to the symmetric equation. Treating $\chi(p)$ as was done before, one can introduce an integral over the energy, make the usual approximations about the phonon energies being much smaller than the Fermi energy and integrate $\mathcal{E}=E+\chi(k)$ analytically (\ref{eq:chi(p)_analytic-integral}), giving\begin{widetext}\begin{align}
    \chi(p)=\;&\frac{1}{2\beta}\matssum{n}\int\frac{d\mathbf{k}}{(2\pi)^3}\gamma_\mathbf{k}V_\text{eff}(p,k)\text{ln}\left[\frac{k_n^2Z^2(k)+\left[\epsilon+\chi(k)\right]^2+W^2(k)}{k_n^2Z^2(k)+\left[\epsilon-\chi(k)\right]^2+W^2(k)}\right]-\frac{1}{2}\int\frac{d\mathbf{q}}{(2\pi)^3}\gamma(\mathbf{p,q,q})(1+2n_B^\mathbf{q}).
\end{align}\end{widetext}
The approximation that $\delta(E-\varepsilon_\mathbf{k})\rightarrow\delta(\epsilon_F-\varepsilon_\mathbf{k})=\gamma_\mathbf{k}$ and $\epsilon\rightarrow+\infty$ relied on the integrand working on the phonon energy scale and the rapid convergence of the integrand in function of $E$. This is not true for the Debye-Waller contribution, which is independent of the electronic band structure. Thus, this approximation cannot be made here and the energy integral is simply done over the introduced Dirac delta distribution $\int dE\delta(E-\varepsilon_\mathbf{k})=1$, equaling unity. $\chi(k)$ works on an energy scale of the phonons, while $\epsilon$ is the maximum energy of the electronic band, assumed to be much larger. Therefore, it made no difference practically if the limit for $\epsilon\rightarrow+\infty$ was taken, which let $\chi(p)\rightarrow0$ self-consistently. Now, there is an additional contribution, providing a lower bound to the value of $\chi(p)$. Thus, letting the boundaries go to infinity still makes the first term approach zero, but the second term remains unaffected, yielding\begin{align}
    \chi(p)=-\frac{1}{2}\int\frac{d\mathbf{q}}{(2\pi)^3}\gamma(\mathbf{p,q,q})(1+2n_B^\mathbf{q})=\Sigma^\text{DW}(\mathbf{p}).
\end{align}
For this to work, $\chi(p)$ of course still has to be negligible to $\epsilon$. Thus, the Debye-Waller self-energy has to be small compared to the maximum energy of the electronic band. This is the additional approximation that has to be made to include the nonlinear electron-phonon coupling. To construct the Eliashberg equations, one has to start by performing the integrals over energy, introduced in the self-energies $S(p)$ and $W(p)$ analytically. Since $\chi(p)$ is non-zero, the energy integral is shifted to new boundaries $-\epsilon+\chi(k)$ and $\epsilon+\chi(k)$. However, its contribution has to be negligible to $\epsilon$ in the first place, ultimately leaving the Eliashberg equations unaffected.

\bibliography{References}

\begin{thebibliography}{60}%
\makeatletter
\providecommand \@ifxundefined [1]{%
 \@ifx{#1\undefined}
}%
\providecommand \@ifnum [1]{%
 \ifnum #1\expandafter \@firstoftwo
 \else \expandafter \@secondoftwo
 \fi
}%
\providecommand \@ifx [1]{%
 \ifx #1\expandafter \@firstoftwo
 \else \expandafter \@secondoftwo
 \fi
}%
\providecommand \natexlab [1]{#1}%
\providecommand \enquote  [1]{``#1''}%
\providecommand \bibnamefont  [1]{#1}%
\providecommand \bibfnamefont [1]{#1}%
\providecommand \citenamefont [1]{#1}%
\providecommand \href@noop [0]{\@secondoftwo}%
\providecommand \href [0]{\begingroup \@sanitize@url \@href}%
\providecommand \@href[1]{\@@startlink{#1}\@@href}%
\providecommand \@@href[1]{\endgroup#1\@@endlink}%
\providecommand \@sanitize@url [0]{\catcode `\\12\catcode `\$12\catcode `\&12\catcode `\#12\catcode `\^12\catcode `\_12\catcode `\%12\relax}%
\providecommand \@@startlink[1]{}%
\providecommand \@@endlink[0]{}%
\providecommand \url  [0]{\begingroup\@sanitize@url \@url }%
\providecommand \@url [1]{\endgroup\@href {#1}{\urlprefix }}%
\providecommand \urlprefix  [0]{URL }%
\providecommand \Eprint [0]{\href }%
\providecommand \doibase [0]{https://doi.org/}%
\providecommand \selectlanguage [0]{\@gobble}%
\providecommand \bibinfo  [0]{\@secondoftwo}%
\providecommand \bibfield  [0]{\@secondoftwo}%
\providecommand \translation [1]{[#1]}%
\providecommand \BibitemOpen [0]{}%
\providecommand \bibitemStop [0]{}%
\providecommand \bibitemNoStop [0]{.\EOS\space}%
\providecommand \EOS [0]{\spacefactor3000\relax}%
\providecommand \BibitemShut  [1]{\csname bibitem#1\endcsname}%
\let\auto@bib@innerbib\@empty
%</preamble>
\bibitem [{\citenamefont {Bardeen}\ \emph {et~al.}(1957)\citenamefont {Bardeen}, \citenamefont {Cooper},\ and\ \citenamefont {Schrieffer}}]{BCS}%
  \BibitemOpen
  \bibfield  {author} {\bibinfo {author} {\bibfnamefont {J.}~\bibnamefont {Bardeen}}, \bibinfo {author} {\bibfnamefont {L.~N.}\ \bibnamefont {Cooper}},\ and\ \bibinfo {author} {\bibfnamefont {J.~R.}\ \bibnamefont {Schrieffer}},\ }\bibfield  {title} {\bibinfo {title} {Theory of superconductivity},\ }\href {https://doi.org/10.1103/PhysRev.108.1175} {\bibfield  {journal} {\bibinfo  {journal} {Phys. Rev.}\ }\textbf {\bibinfo {volume} {108}},\ \bibinfo {pages} {1175} (\bibinfo {year} {1957})}\BibitemShut {NoStop}%
\bibitem [{\citenamefont {Eliashberg}(1960)}]{Eliashberg1960}%
  \BibitemOpen
  \bibfield  {author} {\bibinfo {author} {\bibfnamefont {G.~M.}\ \bibnamefont {Eliashberg}},\ }\bibfield  {title} {\bibinfo {title} {Interactions between electrons and lattice vibrations in a superconductor},\ }\href {http://jetp.ras.ru/cgi-bin/dn/e_011_03_0696.pdf} {\bibfield  {journal} {\bibinfo  {journal} {Sov. Phys. JETP}\ }\textbf {\bibinfo {volume} {11}},\ \bibinfo {pages} {696} (\bibinfo {year} {1960})}\BibitemShut {NoStop}%
\bibitem [{\citenamefont {Eliashberg}(1961)}]{Eliashberg1961}%
  \BibitemOpen
  \bibfield  {author} {\bibinfo {author} {\bibfnamefont {G.~M.}\ \bibnamefont {Eliashberg}},\ }\bibfield  {title} {\bibinfo {title} {Temperature {G}reen’s function for electrons in a superconductor},\ }\href {http://www.jetp.ras.ru/cgi-bin/dn/e_012_05_1000.pdf} {\bibfield  {journal} {\bibinfo  {journal} {Sov. Phys. JETP}\ }\textbf {\bibinfo {volume} {12}},\ \bibinfo {pages} {1000} (\bibinfo {year} {1961})}\BibitemShut {NoStop}%
\bibitem [{\citenamefont {Allen}\ and\ \citenamefont {Mitrovi{\'c}}(1983)}]{allen1983theory}%
  \BibitemOpen
  \bibfield  {author} {\bibinfo {author} {\bibfnamefont {P.~B.}\ \bibnamefont {Allen}}\ and\ \bibinfo {author} {\bibfnamefont {B.}~\bibnamefont {Mitrovi{\'c}}},\ }\bibfield  {title} {\bibinfo {title} {Theory of superconducting {T}c},\ }\href@noop {} {\bibfield  {journal} {\bibinfo  {journal} {Solid state physics}\ }\textbf {\bibinfo {volume} {37}},\ \bibinfo {pages} {1} (\bibinfo {year} {1983})}\BibitemShut {NoStop}%
\bibitem [{\citenamefont {Carbotte}(1990)}]{carbotte1990properties}%
  \BibitemOpen
  \bibfield  {author} {\bibinfo {author} {\bibfnamefont {J.~P.}\ \bibnamefont {Carbotte}},\ }\bibfield  {title} {\bibinfo {title} {Properties of boson-exchange superconductors},\ }\href {https://journals.aps.org/rmp/abstract/10.1103/RevModPhys.62.1027} {\bibfield  {journal} {\bibinfo  {journal} {Reviews of Modern Physics}\ }\textbf {\bibinfo {volume} {62}},\ \bibinfo {pages} {1027} (\bibinfo {year} {1990})}\BibitemShut {NoStop}%
\bibitem [{\citenamefont {Stewart}(2017)}]{Stewart2017}%
  \BibitemOpen
  \bibfield  {author} {\bibinfo {author} {\bibfnamefont {G.~R.}\ \bibnamefont {Stewart}},\ }\bibfield  {title} {\bibinfo {title} {Unconventional superconductivity},\ }\href {https://doi.org/10.1080/00018732.2017.1331615} {\bibfield  {journal} {\bibinfo  {journal} {Advances in Physics}\ }\textbf {\bibinfo {volume} {66}},\ \bibinfo {pages} {75} (\bibinfo {year} {2017})},\ \Eprint {https://arxiv.org/abs/https://doi.org/10.1080/00018732.2017.1331615} {https://doi.org/10.1080/00018732.2017.1331615} \BibitemShut {NoStop}%
\bibitem [{\citenamefont {Maksimov}(2000)}]{Maksimov_2000}%
  \BibitemOpen
  \bibfield  {author} {\bibinfo {author} {\bibfnamefont {E.~G.}\ \bibnamefont {Maksimov}},\ }\bibfield  {title} {\bibinfo {title} {High-temperature superconductivity: the current state},\ }\href {https://doi.org/10.1070/PU2000v043n10ABEH000770} {\bibfield  {journal} {\bibinfo  {journal} {Physics-Uspekhi}\ }\textbf {\bibinfo {volume} {43}},\ \bibinfo {pages} {965} (\bibinfo {year} {2000})}\BibitemShut {NoStop}%
\bibitem [{\citenamefont {Boeri}\ \emph {et~al.}(2022)\citenamefont {Boeri}, \citenamefont {Hennig}, \citenamefont {Hirschfeld}, \citenamefont {Profeta}, \citenamefont {Sanna}, \citenamefont {Zurek}, \citenamefont {Pickett}, \citenamefont {Amsler}, \citenamefont {Dias}, \citenamefont {Eremets}, \citenamefont {Heil}, \citenamefont {Hemley}, \citenamefont {Liu}, \citenamefont {Ma}, \citenamefont {Pierleoni}, \citenamefont {Kolmogorov}, \citenamefont {Rybin}, \citenamefont {Novoselov}, \citenamefont {Anisimov}, \citenamefont {Oganov}, \citenamefont {Pickard}, \citenamefont {Bi}, \citenamefont {Arita}, \citenamefont {Errea}, \citenamefont {Pellegrini}, \citenamefont {Requist}, \citenamefont {Gross}, \citenamefont {Margine}, \citenamefont {Xie}, \citenamefont {Quan}, \citenamefont {Hire}, \citenamefont {Fanfarillo}, \citenamefont {Stewart}, \citenamefont {Hamlin}, \citenamefont {Stanev}, \citenamefont {Gonnelli}, \citenamefont {Piatti}, \citenamefont {Romanin}, \citenamefont {Daghero},\ and\ \citenamefont
  {Valenti}}]{Boeri_2022}%
  \BibitemOpen
  \bibfield  {author} {\bibinfo {author} {\bibfnamefont {L.}~\bibnamefont {Boeri}}, \bibinfo {author} {\bibfnamefont {R.}~\bibnamefont {Hennig}}, \bibinfo {author} {\bibfnamefont {P.}~\bibnamefont {Hirschfeld}}, \bibinfo {author} {\bibfnamefont {G.}~\bibnamefont {Profeta}}, \bibinfo {author} {\bibfnamefont {A.}~\bibnamefont {Sanna}}, \bibinfo {author} {\bibfnamefont {E.}~\bibnamefont {Zurek}}, \bibinfo {author} {\bibfnamefont {W.~E.}\ \bibnamefont {Pickett}}, \bibinfo {author} {\bibfnamefont {M.}~\bibnamefont {Amsler}}, \bibinfo {author} {\bibfnamefont {R.}~\bibnamefont {Dias}}, \bibinfo {author} {\bibfnamefont {M.~I.}\ \bibnamefont {Eremets}}, \bibinfo {author} {\bibfnamefont {C.}~\bibnamefont {Heil}}, \bibinfo {author} {\bibfnamefont {R.~J.}\ \bibnamefont {Hemley}}, \bibinfo {author} {\bibfnamefont {H.}~\bibnamefont {Liu}}, \bibinfo {author} {\bibfnamefont {Y.}~\bibnamefont {Ma}}, \bibinfo {author} {\bibfnamefont {C.}~\bibnamefont {Pierleoni}}, \bibinfo {author} {\bibfnamefont {A.~N.}\ \bibnamefont
  {Kolmogorov}}, \bibinfo {author} {\bibfnamefont {N.}~\bibnamefont {Rybin}}, \bibinfo {author} {\bibfnamefont {D.}~\bibnamefont {Novoselov}}, \bibinfo {author} {\bibfnamefont {V.}~\bibnamefont {Anisimov}}, \bibinfo {author} {\bibfnamefont {A.~R.}\ \bibnamefont {Oganov}}, \bibinfo {author} {\bibfnamefont {C.~J.}\ \bibnamefont {Pickard}}, \bibinfo {author} {\bibfnamefont {T.}~\bibnamefont {Bi}}, \bibinfo {author} {\bibfnamefont {R.}~\bibnamefont {Arita}}, \bibinfo {author} {\bibfnamefont {I.}~\bibnamefont {Errea}}, \bibinfo {author} {\bibfnamefont {C.}~\bibnamefont {Pellegrini}}, \bibinfo {author} {\bibfnamefont {R.}~\bibnamefont {Requist}}, \bibinfo {author} {\bibfnamefont {E.~K.~U.}\ \bibnamefont {Gross}}, \bibinfo {author} {\bibfnamefont {E.~R.}\ \bibnamefont {Margine}}, \bibinfo {author} {\bibfnamefont {S.~R.}\ \bibnamefont {Xie}}, \bibinfo {author} {\bibfnamefont {Y.}~\bibnamefont {Quan}}, \bibinfo {author} {\bibfnamefont {A.}~\bibnamefont {Hire}}, \bibinfo {author} {\bibfnamefont {L.}~\bibnamefont
  {Fanfarillo}}, \bibinfo {author} {\bibfnamefont {G.~R.}\ \bibnamefont {Stewart}}, \bibinfo {author} {\bibfnamefont {J.~J.}\ \bibnamefont {Hamlin}}, \bibinfo {author} {\bibfnamefont {V.}~\bibnamefont {Stanev}}, \bibinfo {author} {\bibfnamefont {R.~S.}\ \bibnamefont {Gonnelli}}, \bibinfo {author} {\bibfnamefont {E.}~\bibnamefont {Piatti}}, \bibinfo {author} {\bibfnamefont {D.}~\bibnamefont {Romanin}}, \bibinfo {author} {\bibfnamefont {D.}~\bibnamefont {Daghero}},\ and\ \bibinfo {author} {\bibfnamefont {R.}~\bibnamefont {Valenti}},\ }\bibfield  {title} {\bibinfo {title} {The 2021 room-temperature superconductivity roadmap},\ }\href {https://doi.org/10.1088/1361-648X/ac2864} {\bibfield  {journal} {\bibinfo  {journal} {Journal of Physics: Condensed Matter}\ }\textbf {\bibinfo {volume} {34}},\ \bibinfo {pages} {183002} (\bibinfo {year} {2022})}\BibitemShut {NoStop}%
\bibitem [{\citenamefont {Kresin}\ \emph {et~al.}(1993)\citenamefont {Kresin}, \citenamefont {Morawitz},\ and\ \citenamefont {Wolf}}]{kresin-1993}%
  \BibitemOpen
  \bibfield  {author} {\bibinfo {author} {\bibfnamefont {V.~Z.}\ \bibnamefont {Kresin}}, \bibinfo {author} {\bibfnamefont {H.}~\bibnamefont {Morawitz}},\ and\ \bibinfo {author} {\bibfnamefont {S.~A.}\ \bibnamefont {Wolf}},\ }\href@noop {} {\emph {\bibinfo {title} {{Mechanisms of conventional and high {T}C superconductivity}}}}\ (\bibinfo  {publisher} {Oxford University Press, USA},\ \bibinfo {year} {1993})\BibitemShut {NoStop}%
\bibitem [{\citenamefont {Lozovik}\ \emph {et~al.}(1976)\citenamefont {Lozovik}, \citenamefont {Yudson},\ and\ \citenamefont {of~Sciences}}]{lozovik-1976}%
  \BibitemOpen
  \bibfield  {author} {\bibinfo {author} {\bibfnamefont {Y.~E.}\ \bibnamefont {Lozovik}}, \bibinfo {author} {\bibfnamefont {V.~I.}\ \bibnamefont {Yudson}},\ and\ \bibinfo {author} {\bibfnamefont {S.~I. U.~A.}\ \bibnamefont {of~Sciences}},\ }\bibfield  {title} {\bibinfo {title} {{A new mechanism for superconductivity: pairing between spatially separated electrons and holes}},\ }\href {http://www.jetp.ras.ru/cgi-bin/dn/e_044_02_0389.pdf} {\bibfield  {journal} {\bibinfo  {journal} {Zh. Eksp. Teor. Fiz.}\ ,\ \bibinfo {pages} {738}} (\bibinfo {year} {1976})}\BibitemShut {NoStop}%
\bibitem [{\citenamefont {Han}\ \emph {et~al.}(2024)\citenamefont {Han}, \citenamefont {Kivelson},\ and\ \citenamefont {Volkov}}]{Han2024}%
  \BibitemOpen
  \bibfield  {author} {\bibinfo {author} {\bibfnamefont {Z.}~\bibnamefont {Han}}, \bibinfo {author} {\bibfnamefont {S.~A.}\ \bibnamefont {Kivelson}},\ and\ \bibinfo {author} {\bibfnamefont {P.~A.}\ \bibnamefont {Volkov}},\ }\bibfield  {title} {\bibinfo {title} {Quantum bipolaron superconductivity from quadratic electron-phonon coupling},\ }\href {https://doi.org/10.1103/PhysRevLett.132.226001} {\bibfield  {journal} {\bibinfo  {journal} {Phys. Rev. Lett.}\ }\textbf {\bibinfo {volume} {132}},\ \bibinfo {pages} {226001} (\bibinfo {year} {2024})}\BibitemShut {NoStop}%
\bibitem [{\citenamefont {Migdal}(1958)}]{Migdal}%
  \BibitemOpen
  \bibfield  {author} {\bibinfo {author} {\bibfnamefont {A.~B.}\ \bibnamefont {Migdal}},\ }\bibfield  {title} {\bibinfo {title} {Interaction between electrons and the lattice vibrations in a normal metal},\ }\href {http://83.149.229.155/cgi-bin/dn/e_007_06_0996.pdf} {\bibfield  {journal} {\bibinfo  {journal} {Zhur. Eksptl'. i Teoret. Fiz.}\ }\textbf {\bibinfo {volume} {34}} (\bibinfo {year} {1958})}\BibitemShut {NoStop}%
\bibitem [{\citenamefont {Schrodi}\ \emph {et~al.}(2021)\citenamefont {Schrodi}, \citenamefont {Aperis},\ and\ \citenamefont {Oppeneer}}]{Schrodi2021}%
  \BibitemOpen
  \bibfield  {author} {\bibinfo {author} {\bibfnamefont {F.}~\bibnamefont {Schrodi}}, \bibinfo {author} {\bibfnamefont {A.}~\bibnamefont {Aperis}},\ and\ \bibinfo {author} {\bibfnamefont {P.~M.}\ \bibnamefont {Oppeneer}},\ }\bibfield  {title} {\bibinfo {title} {Induced odd-frequency superconducting state in vertex-corrected {E}liashberg theory},\ }\href {https://doi.org/10.1103/PhysRevB.104.174518} {\bibfield  {journal} {\bibinfo  {journal} {Phys. Rev. B}\ }\textbf {\bibinfo {volume} {104}},\ \bibinfo {pages} {174518} (\bibinfo {year} {2021})}\BibitemShut {NoStop}%
\bibitem [{\citenamefont {Alexandrov}(2001)}]{Alexandrov_2001}%
  \BibitemOpen
  \bibfield  {author} {\bibinfo {author} {\bibfnamefont {A.~S.}\ \bibnamefont {Alexandrov}},\ }\bibfield  {title} {\bibinfo {title} {Breakdown of the {M}igdal-{E}liashberg theory in the strong-coupling adiabatic regime},\ }\href {https://doi.org/10.1209/epl/i2001-00492-x} {\bibfield  {journal} {\bibinfo  {journal} {Europhysics Letters}\ }\textbf {\bibinfo {volume} {56}},\ \bibinfo {pages} {92} (\bibinfo {year} {2001})}\BibitemShut {NoStop}%
\bibitem [{\citenamefont {Hague}(2007)}]{Hague_2007}%
  \BibitemOpen
  \bibfield  {author} {\bibinfo {author} {\bibfnamefont {J.~P.}\ \bibnamefont {Hague}},\ }\bibfield  {title} {\bibinfo {title} {Failure of conventional superconductivity theory for optical-phonon mediated d-wave pairing},\ }\href {https://doi.org/10.1088/1742-6596/92/1/012119} {\bibfield  {journal} {\bibinfo  {journal} {Journal of Physics: Conference Series}\ }\textbf {\bibinfo {volume} {92}},\ \bibinfo {pages} {012119} (\bibinfo {year} {2007})}\BibitemShut {NoStop}%
\bibitem [{\citenamefont {Hague}\ and\ \citenamefont {d’Ambrumenil}(2008)}]{hague-2008}%
  \BibitemOpen
  \bibfield  {author} {\bibinfo {author} {\bibfnamefont {J.~P.}\ \bibnamefont {Hague}}\ and\ \bibinfo {author} {\bibfnamefont {N.}~\bibnamefont {d’Ambrumenil}},\ }\bibfield  {title} {\bibinfo {title} {{Breakdown of {M}igdal–{E}liashberg theory via catastrophic vertex divergence at low phonon frequency}},\ }\href {https://doi.org/10.1007/s10909-008-9800-z} {\bibfield  {journal} {\bibinfo  {journal} {Journal of Low Temperature Physics}\ }\textbf {\bibinfo {volume} {151}},\ \bibinfo {pages} {1149} (\bibinfo {year} {2008})}\BibitemShut {NoStop}%
\bibitem [{\citenamefont {Miller}\ \emph {et~al.}(1998)\citenamefont {Miller}, \citenamefont {Freericks},\ and\ \citenamefont {Nicol}}]{Miller1998}%
  \BibitemOpen
  \bibfield  {author} {\bibinfo {author} {\bibfnamefont {P.}~\bibnamefont {Miller}}, \bibinfo {author} {\bibfnamefont {J.~K.}\ \bibnamefont {Freericks}},\ and\ \bibinfo {author} {\bibfnamefont {E.~J.}\ \bibnamefont {Nicol}},\ }\bibfield  {title} {\bibinfo {title} {Possible experimentally observable effects of vertex corrections in superconductors},\ }\href {https://doi.org/10.1103/PhysRevB.58.14498} {\bibfield  {journal} {\bibinfo  {journal} {Phys. Rev. B}\ }\textbf {\bibinfo {volume} {58}},\ \bibinfo {pages} {14498} (\bibinfo {year} {1998})}\BibitemShut {NoStop}%
\bibitem [{\citenamefont {Durajski}\ and\ \citenamefont {Szcz{\'e}sniak}(2016)}]{Durajski2016}%
  \BibitemOpen
  \bibfield  {author} {\bibinfo {author} {\bibfnamefont {A.~P.}\ \bibnamefont {Durajski}}\ and\ \bibinfo {author} {\bibfnamefont {R.}~\bibnamefont {Szcz{\'e}sniak}},\ }\bibfield  {title} {\bibinfo {title} {Migdal-{E}liashberg equations - the effective model for superconducting state in {H}$_3${S}},\ }\href {https://arxiv.org/abs/1609.06079} {\bibfield  {journal} {\bibinfo  {journal} {arXiv preprint arXiv:1609.06079}\ } (\bibinfo {year} {2016})}\BibitemShut {NoStop}%
\bibitem [{\citenamefont {Mahan}\ and\ \citenamefont {Sofo}(1993)}]{Mahan1993}%
  \BibitemOpen
  \bibfield  {author} {\bibinfo {author} {\bibfnamefont {G.~D.}\ \bibnamefont {Mahan}}\ and\ \bibinfo {author} {\bibfnamefont {J.~O.}\ \bibnamefont {Sofo}},\ }\bibfield  {title} {\bibinfo {title} {Resistivity and superconductivity from anharmonic phonons},\ }\href {https://doi.org/10.1103/PhysRevB.47.8050} {\bibfield  {journal} {\bibinfo  {journal} {Phys. Rev. B}\ }\textbf {\bibinfo {volume} {47}},\ \bibinfo {pages} {8050} (\bibinfo {year} {1993})}\BibitemShut {NoStop}%
\bibitem [{\citenamefont {Crespi}\ and\ \citenamefont {Cohen}(1993)}]{Crespi-Cohen1993}%
  \BibitemOpen
  \bibfield  {author} {\bibinfo {author} {\bibfnamefont {V.~H.}\ \bibnamefont {Crespi}}\ and\ \bibinfo {author} {\bibfnamefont {M.~L.}\ \bibnamefont {Cohen}},\ }\bibfield  {title} {\bibinfo {title} {Anharmonic phonons and high-temperature superconductivity},\ }\href {https://doi.org/10.1103/PhysRevB.48.398} {\bibfield  {journal} {\bibinfo  {journal} {Phys. Rev. B}\ }\textbf {\bibinfo {volume} {48}},\ \bibinfo {pages} {398} (\bibinfo {year} {1993})}\BibitemShut {NoStop}%
\bibitem [{\citenamefont {Errea}\ \emph {et~al.}(2013)\citenamefont {Errea}, \citenamefont {Calandra},\ and\ \citenamefont {Mauri}}]{Errea2013}%
  \BibitemOpen
  \bibfield  {author} {\bibinfo {author} {\bibfnamefont {I.}~\bibnamefont {Errea}}, \bibinfo {author} {\bibfnamefont {M.}~\bibnamefont {Calandra}},\ and\ \bibinfo {author} {\bibfnamefont {F.}~\bibnamefont {Mauri}},\ }\bibfield  {title} {\bibinfo {title} {First-principles theory of anharmonicity and the inverse isotope effect in superconducting palladium-hydride compounds},\ }\href {https://doi.org/10.1103/PhysRevLett.111.177002} {\bibfield  {journal} {\bibinfo  {journal} {Phys. Rev. Lett.}\ }\textbf {\bibinfo {volume} {111}},\ \bibinfo {pages} {177002} (\bibinfo {year} {2013})}\BibitemShut {NoStop}%
\bibitem [{\citenamefont {Errea}\ \emph {et~al.}(2014)\citenamefont {Errea}, \citenamefont {Calandra},\ and\ \citenamefont {Mauri}}]{Errea2014}%
  \BibitemOpen
  \bibfield  {author} {\bibinfo {author} {\bibfnamefont {I.}~\bibnamefont {Errea}}, \bibinfo {author} {\bibfnamefont {M.}~\bibnamefont {Calandra}},\ and\ \bibinfo {author} {\bibfnamefont {F.}~\bibnamefont {Mauri}},\ }\bibfield  {title} {\bibinfo {title} {Anharmonic free energies and phonon dispersions from the stochastic self-consistent harmonic approximation: Application to platinum and palladium hydrides},\ }\href {https://doi.org/10.1103/PhysRevB.89.064302} {\bibfield  {journal} {\bibinfo  {journal} {Phys. Rev. B}\ }\textbf {\bibinfo {volume} {89}},\ \bibinfo {pages} {064302} (\bibinfo {year} {2014})}\BibitemShut {NoStop}%
\bibitem [{\citenamefont {Errea}\ \emph {et~al.}(2015)\citenamefont {Errea}, \citenamefont {Calandra}, \citenamefont {Pickard}, \citenamefont {Nelson}, \citenamefont {Needs}, \citenamefont {Li}, \citenamefont {Liu}, \citenamefont {Zhang}, \citenamefont {Ma},\ and\ \citenamefont {Mauri}}]{Errea2015}%
  \BibitemOpen
  \bibfield  {author} {\bibinfo {author} {\bibfnamefont {I.}~\bibnamefont {Errea}}, \bibinfo {author} {\bibfnamefont {M.}~\bibnamefont {Calandra}}, \bibinfo {author} {\bibfnamefont {C.~J.}\ \bibnamefont {Pickard}}, \bibinfo {author} {\bibfnamefont {J.}~\bibnamefont {Nelson}}, \bibinfo {author} {\bibfnamefont {R.~J.}\ \bibnamefont {Needs}}, \bibinfo {author} {\bibfnamefont {Y.}~\bibnamefont {Li}}, \bibinfo {author} {\bibfnamefont {H.}~\bibnamefont {Liu}}, \bibinfo {author} {\bibfnamefont {Y.}~\bibnamefont {Zhang}}, \bibinfo {author} {\bibfnamefont {Y.}~\bibnamefont {Ma}},\ and\ \bibinfo {author} {\bibfnamefont {F.}~\bibnamefont {Mauri}},\ }\bibfield  {title} {\bibinfo {title} {High-pressure hydrogen sulfide from first principles: A strongly anharmonic phonon-mediated superconductor},\ }\href {https://doi.org/10.1103/PhysRevLett.114.157004} {\bibfield  {journal} {\bibinfo  {journal} {Phys. Rev. Lett.}\ }\textbf {\bibinfo {volume} {114}},\ \bibinfo {pages} {157004} (\bibinfo {year} {2015})}\BibitemShut {NoStop}%
\bibitem [{\citenamefont {Setty}\ \emph {et~al.}(2021)\citenamefont {Setty}, \citenamefont {Baggioli},\ and\ \citenamefont {Zaccone}}]{Setty2021}%
  \BibitemOpen
  \bibfield  {author} {\bibinfo {author} {\bibfnamefont {C.}~\bibnamefont {Setty}}, \bibinfo {author} {\bibfnamefont {M.}~\bibnamefont {Baggioli}},\ and\ \bibinfo {author} {\bibfnamefont {A.}~\bibnamefont {Zaccone}},\ }\bibfield  {title} {\bibinfo {title} {Anharmonic theory of superconductivity in the high-pressure materials},\ }\href {https://doi.org/10.1103/PhysRevB.103.094519} {\bibfield  {journal} {\bibinfo  {journal} {Phys. Rev. B}\ }\textbf {\bibinfo {volume} {103}},\ \bibinfo {pages} {094519} (\bibinfo {year} {2021})}\BibitemShut {NoStop}%
\bibitem [{\citenamefont {Setty}\ \emph {et~al.}(2024)\citenamefont {Setty}, \citenamefont {Baggioli},\ and\ \citenamefont {Zaccone}}]{Setty2024anharmonic}%
  \BibitemOpen
  \bibfield  {author} {\bibinfo {author} {\bibfnamefont {C.}~\bibnamefont {Setty}}, \bibinfo {author} {\bibfnamefont {M.}~\bibnamefont {Baggioli}},\ and\ \bibinfo {author} {\bibfnamefont {A.}~\bibnamefont {Zaccone}},\ }\bibfield  {title} {\bibinfo {title} {Anharmonic theory of superconductivity and its applications to emerging quantum materials},\ }\href {https://iopscience.iop.org/article/10.1088/1361-648X/ad2159/meta} {\bibfield  {journal} {\bibinfo  {journal} {Journal of Physics: Condensed Matter}\ }\textbf {\bibinfo {volume} {36}},\ \bibinfo {pages} {173002} (\bibinfo {year} {2024})}\BibitemShut {NoStop}%
\bibitem [{\citenamefont {Galbaatar}\ \emph {et~al.}(1991)\citenamefont {Galbaatar}, \citenamefont {Drechsler}, \citenamefont {Plakida},\ and\ \citenamefont {Vuji{\v{c}}i'c}}]{Galbaatar1991isotope}%
  \BibitemOpen
  \bibfield  {author} {\bibinfo {author} {\bibfnamefont {T.}~\bibnamefont {Galbaatar}}, \bibinfo {author} {\bibfnamefont {S.~L.}\ \bibnamefont {Drechsler}}, \bibinfo {author} {\bibfnamefont {N.~M.}\ \bibnamefont {Plakida}},\ and\ \bibinfo {author} {\bibfnamefont {G.~M.}\ \bibnamefont {Vuji{\v{c}}i'c}},\ }\bibfield  {title} {\bibinfo {title} {Isotope effect in a superconductor with strong anharmonicity},\ }\href {https://www.sciencedirect.com/science/article/pii/0921453491900543} {\bibfield  {journal} {\bibinfo  {journal} {Physica C: Superconductivity}\ }\textbf {\bibinfo {volume} {176}},\ \bibinfo {pages} {496} (\bibinfo {year} {1991})}\BibitemShut {NoStop}%
\bibitem [{\citenamefont {Lucrezi}\ \emph {et~al.}(2024)\citenamefont {Lucrezi}, \citenamefont {Ferreira}, \citenamefont {Aichhorn},\ and\ \citenamefont {Heil}}]{Lucrezi2023}%
  \BibitemOpen
  \bibfield  {author} {\bibinfo {author} {\bibfnamefont {R.}~\bibnamefont {Lucrezi}}, \bibinfo {author} {\bibfnamefont {P.~P.}\ \bibnamefont {Ferreira}}, \bibinfo {author} {\bibfnamefont {M.}~\bibnamefont {Aichhorn}},\ and\ \bibinfo {author} {\bibfnamefont {C.}~\bibnamefont {Heil}},\ }\bibfield  {title} {\bibinfo {title} {Temperature and quantum anharmonic lattice effects on stability and superconductivity in lutetium trihydride},\ }\bibfield  {journal} {\bibinfo  {journal} {Nature Communications}\ }\textbf {\bibinfo {volume} {15}},\ \href {https://doi.org/10.1038/s41467-023-44326-4} {10.1038/s41467-023-44326-4} (\bibinfo {year} {2024})\BibitemShut {NoStop}%
\bibitem [{\citenamefont {Choi}\ \emph {et~al.}(2002)\citenamefont {Choi}, \citenamefont {Roundy}, \citenamefont {Sun}, \citenamefont {Cohen},\ and\ \citenamefont {Louie}}]{Choi}%
  \BibitemOpen
  \bibfield  {author} {\bibinfo {author} {\bibfnamefont {H.~J.}\ \bibnamefont {Choi}}, \bibinfo {author} {\bibfnamefont {D.}~\bibnamefont {Roundy}}, \bibinfo {author} {\bibfnamefont {H.}~\bibnamefont {Sun}}, \bibinfo {author} {\bibfnamefont {M.~L.}\ \bibnamefont {Cohen}},\ and\ \bibinfo {author} {\bibfnamefont {S.~G.}\ \bibnamefont {Louie}},\ }\bibfield  {title} {\bibinfo {title} {First-principles calculation of the superconducting transition in {M}g{B}$_{2}$ within the anisotropic {E}liashberg formalism},\ }\href {https://doi.org/10.1103/PhysRevB.66.020513} {\bibfield  {journal} {\bibinfo  {journal} {Phys. Rev. B}\ }\textbf {\bibinfo {volume} {66}},\ \bibinfo {pages} {020513} (\bibinfo {year} {2002})}\BibitemShut {NoStop}%
\bibitem [{\citenamefont {Bianco}\ and\ \citenamefont {Errea}(2023)}]{bianco2023}%
  \BibitemOpen
  \bibfield  {author} {\bibinfo {author} {\bibfnamefont {R.}~\bibnamefont {Bianco}}\ and\ \bibinfo {author} {\bibfnamefont {I.}~\bibnamefont {Errea}},\ }\bibfield  {title} {\bibinfo {title} {Non-perturbative theory of the electron-phonon coupling and its first-principles implementation},\ }\href {https://arxiv.org/abs/2303.02621} {\bibfield  {journal} {\bibinfo  {journal} {arXiv preprint arXiv:2303.02621}\ } (\bibinfo {year} {2023})}\BibitemShut {NoStop}%
\bibitem [{\citenamefont {Liu}\ \emph {et~al.}(2001)\citenamefont {Liu}, \citenamefont {Mazin},\ and\ \citenamefont {Kortus}}]{Liu2001}%
  \BibitemOpen
  \bibfield  {author} {\bibinfo {author} {\bibfnamefont {A.~Y.}\ \bibnamefont {Liu}}, \bibinfo {author} {\bibfnamefont {I.~I.}\ \bibnamefont {Mazin}},\ and\ \bibinfo {author} {\bibfnamefont {J.}~\bibnamefont {Kortus}},\ }\bibfield  {title} {\bibinfo {title} {Beyond {E}liashberg superconductivity in {M}g{B}$_2$: Anharmonicity, two-phonon scattering, and multiple gaps},\ }\href {https://doi.org/10.1103/PhysRevLett.87.087005} {\bibfield  {journal} {\bibinfo  {journal} {Phys. Rev. Lett.}\ }\textbf {\bibinfo {volume} {87}},\ \bibinfo {pages} {087005} (\bibinfo {year} {2001})}\BibitemShut {NoStop}%
\bibitem [{\citenamefont {Adolphs}\ and\ \citenamefont {Berciu}(2013)}]{Adolphs_2013}%
  \BibitemOpen
  \bibfield  {author} {\bibinfo {author} {\bibfnamefont {C.~P.~J.}\ \bibnamefont {Adolphs}}\ and\ \bibinfo {author} {\bibfnamefont {M.}~\bibnamefont {Berciu}},\ }\bibfield  {title} {\bibinfo {title} {Going beyond the linear approximation in describing electron-phonon coupling: Relevance for the {H}olstein model},\ }\href {https://doi.org/10.1209/0295-5075/102/47003} {\bibfield  {journal} {\bibinfo  {journal} {EPL (Europhysics Letters)}\ }\textbf {\bibinfo {volume} {102}},\ \bibinfo {pages} {47003} (\bibinfo {year} {2013})}\BibitemShut {NoStop}%
\bibitem [{\citenamefont {Chubukov}\ \emph {et~al.}(2020)\citenamefont {Chubukov}, \citenamefont {Abanov}, \citenamefont {Esterlis},\ and\ \citenamefont {Kivelson}}]{chubukov2020eliashberg}%
  \BibitemOpen
  \bibfield  {author} {\bibinfo {author} {\bibfnamefont {A.~V.}\ \bibnamefont {Chubukov}}, \bibinfo {author} {\bibfnamefont {A.}~\bibnamefont {Abanov}}, \bibinfo {author} {\bibfnamefont {I.}~\bibnamefont {Esterlis}},\ and\ \bibinfo {author} {\bibfnamefont {S.~A.}\ \bibnamefont {Kivelson}},\ }\bibfield  {title} {\bibinfo {title} {Eliashberg theory of phonon-mediated superconductivity—when it is valid and how it breaks down},\ }\href {https://www.sciencedirect.com/science/article/pii/S0003491620301238} {\bibfield  {journal} {\bibinfo  {journal} {Annals of Physics}\ }\textbf {\bibinfo {volume} {417}},\ \bibinfo {pages} {168190} (\bibinfo {year} {2020})}\BibitemShut {NoStop}%
\bibitem [{\citenamefont {Dynes}\ \emph {et~al.}(1986)\citenamefont {Dynes}, \citenamefont {White}, \citenamefont {Graybeal},\ and\ \citenamefont {Garno}}]{Dynes1986}%
  \BibitemOpen
  \bibfield  {author} {\bibinfo {author} {\bibfnamefont {R.~C.}\ \bibnamefont {Dynes}}, \bibinfo {author} {\bibfnamefont {A.~E.}\ \bibnamefont {White}}, \bibinfo {author} {\bibfnamefont {J.~M.}\ \bibnamefont {Graybeal}},\ and\ \bibinfo {author} {\bibfnamefont {J.~P.}\ \bibnamefont {Garno}},\ }\bibfield  {title} {\bibinfo {title} {Breakdown of {E}liashberg theory for two-dimensional superconductivity in the presence of disorder},\ }\href {https://doi.org/10.1103/PhysRevLett.57.2195} {\bibfield  {journal} {\bibinfo  {journal} {Phys. Rev. Lett.}\ }\textbf {\bibinfo {volume} {57}},\ \bibinfo {pages} {2195} (\bibinfo {year} {1986})}\BibitemShut {NoStop}%
\bibitem [{\citenamefont {Tikhonov}\ \emph {et~al.}(2020)\citenamefont {Tikhonov}, \citenamefont {Semenov}, \citenamefont {Devyatov},\ and\ \citenamefont {Skvortsov}}]{tikhonov2020microwave}%
  \BibitemOpen
  \bibfield  {author} {\bibinfo {author} {\bibfnamefont {K.~S.}\ \bibnamefont {Tikhonov}}, \bibinfo {author} {\bibfnamefont {A.~V.}\ \bibnamefont {Semenov}}, \bibinfo {author} {\bibfnamefont {I.~A.}\ \bibnamefont {Devyatov}},\ and\ \bibinfo {author} {\bibfnamefont {M.~A.}\ \bibnamefont {Skvortsov}},\ }\bibfield  {title} {\bibinfo {title} {Microwave response of a superconductor beyond the {E}liashberg theory},\ }\href {https://www.sciencedirect.com/science/article/pii/S0003491620300348} {\bibfield  {journal} {\bibinfo  {journal} {Annals of Physics}\ }\textbf {\bibinfo {volume} {417}},\ \bibinfo {pages} {168101} (\bibinfo {year} {2020})}\BibitemShut {NoStop}%
\bibitem [{\citenamefont {van~der Marel}\ \emph {et~al.}(2019)\citenamefont {van~der Marel}, \citenamefont {Barantani},\ and\ \citenamefont {Rischau}}]{vanderMarel2019}%
  \BibitemOpen
  \bibfield  {author} {\bibinfo {author} {\bibfnamefont {D.}~\bibnamefont {van~der Marel}}, \bibinfo {author} {\bibfnamefont {F.}~\bibnamefont {Barantani}},\ and\ \bibinfo {author} {\bibfnamefont {C.~W.}\ \bibnamefont {Rischau}},\ }\bibfield  {title} {\bibinfo {title} {Possible mechanism for superconductivity in doped {S}r{T}i{O}$_{3}$},\ }\href {https://doi.org/10.1103/PhysRevResearch.1.013003} {\bibfield  {journal} {\bibinfo  {journal} {Phys. Rev. Res.}\ }\textbf {\bibinfo {volume} {1}},\ \bibinfo {pages} {013003} (\bibinfo {year} {2019})}\BibitemShut {NoStop}%
\bibitem [{\citenamefont {Collignon}\ \emph {et~al.}(2019)\citenamefont {Collignon}, \citenamefont {Lin}, \citenamefont {Rischau}, \citenamefont {Fauqué},\ and\ \citenamefont {Behnia}}]{Collignon2019}%
  \BibitemOpen
  \bibfield  {author} {\bibinfo {author} {\bibfnamefont {C.}~\bibnamefont {Collignon}}, \bibinfo {author} {\bibfnamefont {X.}~\bibnamefont {Lin}}, \bibinfo {author} {\bibfnamefont {C.~W.}\ \bibnamefont {Rischau}}, \bibinfo {author} {\bibfnamefont {B.}~\bibnamefont {Fauqué}},\ and\ \bibinfo {author} {\bibfnamefont {K.}~\bibnamefont {Behnia}},\ }\bibfield  {title} {\bibinfo {title} {Metallicity and superconductivity in doped strontium titanate},\ }\href {https://doi.org/https://doi.org/10.1146/annurev-conmatphys-031218-013144} {\bibfield  {journal} {\bibinfo  {journal} {Annual Review of Condensed Matter Physics}\ }\textbf {\bibinfo {volume} {10}},\ \bibinfo {pages} {25} (\bibinfo {year} {2019})}\BibitemShut {NoStop}%
\bibitem [{\citenamefont {Gastiasoro}\ \emph {et~al.}(2020)\citenamefont {Gastiasoro}, \citenamefont {Ruhman},\ and\ \citenamefont {Fernandes}}]{GASTIASORO2020}%
  \BibitemOpen
  \bibfield  {author} {\bibinfo {author} {\bibfnamefont {M.~N.}\ \bibnamefont {Gastiasoro}}, \bibinfo {author} {\bibfnamefont {J.}~\bibnamefont {Ruhman}},\ and\ \bibinfo {author} {\bibfnamefont {R.~M.}\ \bibnamefont {Fernandes}},\ }\bibfield  {title} {\bibinfo {title} {Superconductivity in dilute {S}r{T}io3: A review},\ }\href {https://doi.org/https://doi.org/10.1016/j.aop.2020.168107} {\bibfield  {journal} {\bibinfo  {journal} {Annals of Physics}\ }\textbf {\bibinfo {volume} {417}},\ \bibinfo {pages} {168107} (\bibinfo {year} {2020})}\BibitemShut {NoStop}%
\bibitem [{\citenamefont {Protter}\ \emph {et~al.}(2021)\citenamefont {Protter}, \citenamefont {Boyack},\ and\ \citenamefont {Marsiglio}}]{Protter2021}%
  \BibitemOpen
  \bibfield  {author} {\bibinfo {author} {\bibfnamefont {M.}~\bibnamefont {Protter}}, \bibinfo {author} {\bibfnamefont {R.}~\bibnamefont {Boyack}},\ and\ \bibinfo {author} {\bibfnamefont {F.}~\bibnamefont {Marsiglio}},\ }\bibfield  {title} {\bibinfo {title} {Functional-integral approach to gaussian fluctuations in eliashberg theory},\ }\href {https://doi.org/10.1103/PhysRevB.104.014513} {\bibfield  {journal} {\bibinfo  {journal} {Phys. Rev. B}\ }\textbf {\bibinfo {volume} {104}},\ \bibinfo {pages} {014513} (\bibinfo {year} {2021})}\BibitemShut {NoStop}%
\bibitem [{\citenamefont {Dalal}\ \emph {et~al.}(2023)\citenamefont {Dalal}, \citenamefont {Ruhman},\ and\ \citenamefont {Kozii}}]{Dalal2023}%
  \BibitemOpen
  \bibfield  {author} {\bibinfo {author} {\bibfnamefont {A.}~\bibnamefont {Dalal}}, \bibinfo {author} {\bibfnamefont {J.}~\bibnamefont {Ruhman}},\ and\ \bibinfo {author} {\bibfnamefont {V.}~\bibnamefont {Kozii}},\ }\bibfield  {title} {\bibinfo {title} {Field theory of a superconductor with repulsion},\ }\href {https://doi.org/10.1103/PhysRevB.108.214521} {\bibfield  {journal} {\bibinfo  {journal} {Phys. Rev. B}\ }\textbf {\bibinfo {volume} {108}},\ \bibinfo {pages} {214521} (\bibinfo {year} {2023})}\BibitemShut {NoStop}%
\bibitem [{\citenamefont {Aase}\ \emph {et~al.}(2023)\citenamefont {Aase}, \citenamefont {M\ae{}land},\ and\ \citenamefont {Sudb\o{}}}]{Aase2023}%
  \BibitemOpen
  \bibfield  {author} {\bibinfo {author} {\bibfnamefont {N.~H.}\ \bibnamefont {Aase}}, \bibinfo {author} {\bibfnamefont {K.}~\bibnamefont {M\ae{}land}},\ and\ \bibinfo {author} {\bibfnamefont {A.}~\bibnamefont {Sudb\o{}}},\ }\bibfield  {title} {\bibinfo {title} {Multiband strong-coupling superconductors with spontaneously broken time-reversal symmetry},\ }\href {https://doi.org/10.1103/PhysRevB.108.214508} {\bibfield  {journal} {\bibinfo  {journal} {Phys. Rev. B}\ }\textbf {\bibinfo {volume} {108}},\ \bibinfo {pages} {214508} (\bibinfo {year} {2023})}\BibitemShut {NoStop}%
\bibitem [{\citenamefont {Marsiglio}(2020)}]{Marsiglio}%
  \BibitemOpen
  \bibfield  {author} {\bibinfo {author} {\bibfnamefont {F.}~\bibnamefont {Marsiglio}},\ }\bibfield  {title} {\bibinfo {title} {Eliashberg theory: A short review},\ }\href {https://doi.org/10.1016/j.aop.2020.168102} {\bibfield  {journal} {\bibinfo  {journal} {Annals of Physics}\ }\textbf {\bibinfo {volume} {417}},\ \bibinfo {pages} {168102} (\bibinfo {year} {2020})}\BibitemShut {NoStop}%
\bibitem [{\citenamefont {Karakozov}\ and\ \citenamefont {Maksimov}(1978)}]{Karakozov-Maksimov}%
  \BibitemOpen
  \bibfield  {author} {\bibinfo {author} {\bibfnamefont {A.~E.}\ \bibnamefont {Karakozov}}\ and\ \bibinfo {author} {\bibfnamefont {E.~G.}\ \bibnamefont {Maksimov}},\ }\bibfield  {title} {\bibinfo {title} {Influence of anharmonicity on superconductivity},\ }\href {http://www.jetp.ras.ru/cgi-bin/dn/e_047_02_0358.pdf} {\bibfield  {journal} {\bibinfo  {journal} {Zh. Eskp. Teor. Fiz}\ }\textbf {\bibinfo {volume} {74}},\ \bibinfo {pages} {681} (\bibinfo {year} {1978})}\BibitemShut {NoStop}%
\bibitem [{\citenamefont {Mahan}(2010)}]{Mahan}%
  \BibitemOpen
  \bibfield  {author} {\bibinfo {author} {\bibfnamefont {G.~D.}\ \bibnamefont {Mahan}},\ }\href@noop {} {\emph {\bibinfo {title} {{Many-Particle Physics}}}}\ (\bibinfo  {publisher} {Springer},\ \bibinfo {year} {2010})\BibitemShut {NoStop}%
\bibitem [{\citenamefont {Marsiglio}\ and\ \citenamefont {Carbotte}(2008)}]{marsiglio2001}%
  \BibitemOpen
  \bibfield  {author} {\bibinfo {author} {\bibfnamefont {F.}~\bibnamefont {Marsiglio}}\ and\ \bibinfo {author} {\bibfnamefont {J.~P.}\ \bibnamefont {Carbotte}},\ }\href@noop {} {\bibinfo {title} {Electron-phonon superconductivity}} (\bibinfo {year} {2008})\BibitemShut {NoStop}%
\bibitem [{\citenamefont {Pavarini}\ \emph {et~al.}(2013)\citenamefont {Pavarini}, \citenamefont {Koch},\ and\ \citenamefont {Schollwöck}}]{pavarini-2013}%
  \BibitemOpen
  \bibfield  {author} {\bibinfo {author} {\bibfnamefont {E.}~\bibnamefont {Pavarini}}, \bibinfo {author} {\bibfnamefont {E.}~\bibnamefont {Koch}},\ and\ \bibinfo {author} {\bibfnamefont {U.}~\bibnamefont {Schollwöck}},\ }\href@noop {} {\emph {\bibinfo {title} {{Emergent phenomena in correlated matter}}}}\ (\bibinfo  {publisher} {Forschungszentrum Jülich},\ \bibinfo {year} {2013})\BibitemShut {NoStop}%
\bibitem [{\citenamefont {Monacelli}\ \emph {et~al.}(2021)\citenamefont {Monacelli}, \citenamefont {Bianco}, \citenamefont {Cherubini}, \citenamefont {Calandra}, \citenamefont {Errea},\ and\ \citenamefont {Mauri}}]{Monacelli_2021}%
  \BibitemOpen
  \bibfield  {author} {\bibinfo {author} {\bibfnamefont {L.}~\bibnamefont {Monacelli}}, \bibinfo {author} {\bibfnamefont {R.}~\bibnamefont {Bianco}}, \bibinfo {author} {\bibfnamefont {M.}~\bibnamefont {Cherubini}}, \bibinfo {author} {\bibfnamefont {M.}~\bibnamefont {Calandra}}, \bibinfo {author} {\bibfnamefont {I.}~\bibnamefont {Errea}},\ and\ \bibinfo {author} {\bibfnamefont {F.}~\bibnamefont {Mauri}},\ }\bibfield  {title} {\bibinfo {title} {The stochastic self-consistent harmonic approximation: calculating vibrational properties of materials with full quantum and anharmonic effects},\ }\href {https://doi.org/10.1088/1361-648X/ac066b} {\bibfield  {journal} {\bibinfo  {journal} {Journal of Physics: Condensed Matter}\ }\textbf {\bibinfo {volume} {33}},\ \bibinfo {pages} {363001} (\bibinfo {year} {2021})}\BibitemShut {NoStop}%
\bibitem [{\citenamefont {Gor’Kov}(1958)}]{Gor'kov}%
  \BibitemOpen
  \bibfield  {author} {\bibinfo {author} {\bibfnamefont {L.~P.}\ \bibnamefont {Gor’Kov}},\ }\bibfield  {title} {\bibinfo {title} {On the energy spectrum of superconductors},\ }\href {http://jetp.ras.ru/cgi-bin/dn/e_007_03_0505.pdf} {\bibfield  {journal} {\bibinfo  {journal} {Sov. Phys. JETP}\ }\textbf {\bibinfo {volume} {7}},\ \bibinfo {pages} {158} (\bibinfo {year} {1958})}\BibitemShut {NoStop}%
\bibitem [{\citenamefont {Allen}(1976)}]{Allen}%
  \BibitemOpen
  \bibfield  {author} {\bibinfo {author} {\bibfnamefont {P.~B.}\ \bibnamefont {Allen}},\ }\bibfield  {title} {\bibinfo {title} {Fermi-surface harmonics: A general method for nonspherical problems. application to {B}oltzmann and {E}liashberg equations},\ }\href {https://doi.org/10.1103/PhysRevB.13.1416} {\bibfield  {journal} {\bibinfo  {journal} {Phys. Rev. B}\ }\textbf {\bibinfo {volume} {13}},\ \bibinfo {pages} {1416} (\bibinfo {year} {1976})}\BibitemShut {NoStop}%
\bibitem [{\citenamefont {Margine}\ and\ \citenamefont {Giustino}(2013)}]{Margine_Giustino}%
  \BibitemOpen
  \bibfield  {author} {\bibinfo {author} {\bibfnamefont {E.~R.}\ \bibnamefont {Margine}}\ and\ \bibinfo {author} {\bibfnamefont {F.}~\bibnamefont {Giustino}},\ }\bibfield  {title} {\bibinfo {title} {Anisotropic {M}igdal-{E}liashberg theory using {W}annier functions},\ }\href {https://doi.org/10.1103/PhysRevB.87.024505} {\bibfield  {journal} {\bibinfo  {journal} {Phys. Rev. B}\ }\textbf {\bibinfo {volume} {87}},\ \bibinfo {pages} {024505} (\bibinfo {year} {2013})}\BibitemShut {NoStop}%
\bibitem [{\citenamefont {Davydov}(2018)}]{gross2018coulomb}%
  \BibitemOpen
  \bibfield  {author} {\bibinfo {author} {\bibfnamefont {A.}~\bibnamefont {Davydov}},\ }\emph {\bibinfo {title} {Coulomb interaction in the {E}liashberg theory of Superconductivity}},\ \href {https://repo.bibliothek.uni-halle.de/bitstream/1981185920/8994/1/thesis_ArkadiyDavydov.pdf} {Ph.D. thesis},\ \bibinfo  {school} {TU Graz} (\bibinfo {year} {2018})\BibitemShut {NoStop}%
\bibitem [{\citenamefont {Morel}\ and\ \citenamefont {Anderson}(1962)}]{Morel-Anderson}%
  \BibitemOpen
  \bibfield  {author} {\bibinfo {author} {\bibfnamefont {P.}~\bibnamefont {Morel}}\ and\ \bibinfo {author} {\bibfnamefont {P.~W.}\ \bibnamefont {Anderson}},\ }\bibfield  {title} {\bibinfo {title} {Calculation of the superconducting state parameters with retarded electron-phonon interaction},\ }\href {https://doi.org/10.1103/PhysRev.125.1263} {\bibfield  {journal} {\bibinfo  {journal} {Phys. Rev.}\ }\textbf {\bibinfo {volume} {125}},\ \bibinfo {pages} {1263} (\bibinfo {year} {1962})}\BibitemShut {NoStop}%
\bibitem [{\citenamefont {Giustino}(2017)}]{Giustino2017}%
  \BibitemOpen
  \bibfield  {author} {\bibinfo {author} {\bibfnamefont {F.}~\bibnamefont {Giustino}},\ }\bibfield  {title} {\bibinfo {title} {Electron-phonon interactions from first principles},\ }\href {https://doi.org/10.1103/RevModPhys.89.015003} {\bibfield  {journal} {\bibinfo  {journal} {Rev. Mod. Phys.}\ }\textbf {\bibinfo {volume} {89}},\ \bibinfo {pages} {015003} (\bibinfo {year} {2017})}\BibitemShut {NoStop}%
\bibitem [{\citenamefont {Fr{\"o}hlich}(1954)}]{Frohlich}%
  \BibitemOpen
  \bibfield  {author} {\bibinfo {author} {\bibfnamefont {H.}~\bibnamefont {Fr{\"o}hlich}},\ }\bibfield  {title} {\bibinfo {title} {Electrons in lattice fields},\ }\href@noop {} {\bibfield  {journal} {\bibinfo  {journal} {Advances in Physics}\ }\textbf {\bibinfo {volume} {3}},\ \bibinfo {pages} {325} (\bibinfo {year} {1954})}\BibitemShut {NoStop}%
\bibitem [{\citenamefont {Marsiglio}\ \emph {et~al.}(1988)\citenamefont {Marsiglio}, \citenamefont {Schossmann},\ and\ \citenamefont {Carbotte}}]{Marsiglio_analytical}%
  \BibitemOpen
  \bibfield  {author} {\bibinfo {author} {\bibfnamefont {F.}~\bibnamefont {Marsiglio}}, \bibinfo {author} {\bibfnamefont {M.}~\bibnamefont {Schossmann}},\ and\ \bibinfo {author} {\bibfnamefont {J.~P.}\ \bibnamefont {Carbotte}},\ }\bibfield  {title} {\bibinfo {title} {{Iterative analytic continuation of the electron self-energy to the real axis}},\ }\href {https://doi.org/10.1103/physrevb.37.4965} {\bibfield  {journal} {\bibinfo  {journal} {Physical review. B, Condensed matter}\ }\textbf {\bibinfo {volume} {37}},\ \bibinfo {pages} {4965} (\bibinfo {year} {1988})}\BibitemShut {NoStop}%
\bibitem [{\citenamefont {Houtput}\ and\ \citenamefont {Tempere}(2022)}]{houtput-2022}%
  \BibitemOpen
  \bibfield  {author} {\bibinfo {author} {\bibfnamefont {M.}~\bibnamefont {Houtput}}\ and\ \bibinfo {author} {\bibfnamefont {J.}~\bibnamefont {Tempere}},\ }\bibfield  {title} {\bibinfo {title} {{Optical conductivity of an anharmonic large polaron gas at weak coupling}},\ }\bibfield  {journal} {\bibinfo  {journal} {Physical review. B.}\ }\textbf {\bibinfo {volume} {106}},\ \href {https://doi.org/10.1103/physrevb.106.214315} {10.1103/physrevb.106.214315} (\bibinfo {year} {2022})\BibitemShut {NoStop}%
\bibitem [{\citenamefont {Houtput}\ \emph {et~al.}(2024)\citenamefont {Houtput}, \citenamefont {Ranalli}, \citenamefont {Verdi}, \citenamefont {Klimin}, \citenamefont {Ragni}, \citenamefont {Franchini},\ and\ \citenamefont {Tempere}}]{houtput2024}%
  \BibitemOpen
  \bibfield  {author} {\bibinfo {author} {\bibfnamefont {M.}~\bibnamefont {Houtput}}, \bibinfo {author} {\bibfnamefont {L.}~\bibnamefont {Ranalli}}, \bibinfo {author} {\bibfnamefont {C.}~\bibnamefont {Verdi}}, \bibinfo {author} {\bibfnamefont {S.}~\bibnamefont {Klimin}}, \bibinfo {author} {\bibfnamefont {S.}~\bibnamefont {Ragni}}, \bibinfo {author} {\bibfnamefont {C.}~\bibnamefont {Franchini}},\ and\ \bibinfo {author} {\bibfnamefont {J.}~\bibnamefont {Tempere}},\ }\href {https://arxiv.org/abs/2412.09470} {\bibinfo {title} {Anharmonic long-range electron-phonon interaction: analytic formalism}} (\bibinfo {year} {2024}),\ \Eprint {https://arxiv.org/abs/2412.09470} {arXiv:2412.09470 [cond-mat.mtrl-sci]} \BibitemShut {NoStop}%
\bibitem [{\citenamefont {Leavens}\ and\ \citenamefont {Carbotte}(1971)}]{leavens1971ratio}%
  \BibitemOpen
  \bibfield  {author} {\bibinfo {author} {\bibfnamefont {C.~R.}\ \bibnamefont {Leavens}}\ and\ \bibinfo {author} {\bibfnamefont {J.~P.}\ \bibnamefont {Carbotte}},\ }\bibfield  {title} {\bibinfo {title} {Contribution to the theory of weak coupling superconductors},\ }\href {https://cdnsciencepub.com/doi/abs/10.1139/p71-088} {\bibfield  {journal} {\bibinfo  {journal} {Canadian Journal of Physics}\ }\textbf {\bibinfo {volume} {49}},\ \bibinfo {pages} {724} (\bibinfo {year} {1971})}\BibitemShut {NoStop}%
\bibitem [{\citenamefont {Leavens}(1973)}]{leavens1973ratio}%
  \BibitemOpen
  \bibfield  {author} {\bibinfo {author} {\bibfnamefont {C.~R.}\ \bibnamefont {Leavens}},\ }\bibfield  {title} {\bibinfo {title} {A simple and accurate equation for the superconducting transition temperature},\ }\href {https://www.sciencedirect.com/science/article/pii/0038109873902482} {\bibfield  {journal} {\bibinfo  {journal} {Solid State Communications}\ }\textbf {\bibinfo {volume} {13}},\ \bibinfo {pages} {1607} (\bibinfo {year} {1973})}\BibitemShut {NoStop}%
\bibitem [{\citenamefont {Mitrovi{\'c}}\ \emph {et~al.}(1984)\citenamefont {Mitrovi{\'c}}, \citenamefont {Zarate},\ and\ \citenamefont {Carbotte}}]{mitrovic1984ratio}%
  \BibitemOpen
  \bibfield  {author} {\bibinfo {author} {\bibfnamefont {B.}~\bibnamefont {Mitrovi{\'c}}}, \bibinfo {author} {\bibfnamefont {H.~G.}\ \bibnamefont {Zarate}},\ and\ \bibinfo {author} {\bibfnamefont {J.~P.}\ \bibnamefont {Carbotte}},\ }\bibfield  {title} {\bibinfo {title} {The ratio $2\mathrm{\Delta}_0/k_{B}{T}_c$ within eliashberg theory},\ }\href {https://journals.aps.org/prb/abstract/10.1103/PhysRevB.29.184} {\bibfield  {journal} {\bibinfo  {journal} {Physical Review B}\ }\textbf {\bibinfo {volume} {29}},\ \bibinfo {pages} {184} (\bibinfo {year} {1984})}\BibitemShut {NoStop}%
\bibitem [{\citenamefont {Saidi}\ \emph {et~al.}(2016)\citenamefont {Saidi}, \citenamefont {Ponc{\'e}},\ and\ \citenamefont {Monserrat}}]{Saidi2016}%
  \BibitemOpen
  \bibfield  {author} {\bibinfo {author} {\bibfnamefont {W.~A.}\ \bibnamefont {Saidi}}, \bibinfo {author} {\bibfnamefont {S.}~\bibnamefont {Ponc{\'e}}},\ and\ \bibinfo {author} {\bibfnamefont {B.}~\bibnamefont {Monserrat}},\ }\bibfield  {title} {\bibinfo {title} {Temperature dependence of the energy levels of methylammonium lead iodide perovskite from first-principles},\ }\href {https://doi.org/10.1021/acs.jpclett.6b02560} {\bibfield  {journal} {\bibinfo  {journal} {The Journal of Physical Chemistry Letters}\ }\textbf {\bibinfo {volume} {7}},\ \bibinfo {pages} {5247} (\bibinfo {year} {2016})},\ \bibinfo {note} {pMID: 27973908},\ \Eprint {https://arxiv.org/abs/https://doi.org/10.1021/acs.jpclett.6b02560} {https://doi.org/10.1021/acs.jpclett.6b02560} \BibitemShut {NoStop}%
\end{thebibliography}%
\end{document}